\documentclass[aps,pra,twocolumn,groupedaddress]{revtex4} 
\usepackage{epsf}
\usepackage[dvips]{color}
\usepackage{graphicx}
\usepackage{amsmath} 
\usepackage{pstricks}

\begin{document}
\title{Hall effects in Bose--Einstein condensates in a rotating optical lattice}
\author{Rajiv Bhat, M.~Kr\"amer, J.~Cooper and M.~J.~Holland}

\address{ JILA and Department of Physics, University of Colorado at Boulder, Colorado 80309-0440, USA}

\begin{abstract}
Using the Kubo formalism, we demonstrate fractional quantum Hall features in a rotating Bose-Einstein condensate in a co-rotating two-dimensional optical lattice.
The co-rotating lattice and trap potential allow for an effective magnetic field and compensation of the centrifugal potential. Fractional quantum Hall features are seen for the single-particle system and for few strongly interacting many-particle systems. 
\end{abstract}
\maketitle

\section{Introduction}
Rotating Bose-Einstein condensates (BECs) have been a subject of great theoretical and experimental interest over the last few years. Starting from the quantum engineering of a single vortex~\cite{Williams:1999}, rotating condensates have been used to understand exotic phenomena such as the formation of Abrikosov vortex lattices~\cite{Madison:2000, Haljan:2001} and the BCS-BEC crossover~\cite{Zwierlein:2005}. Parallels have been drawn between a rotating condensate and electrons in the presence of a magnetic field and, in particular, the fractional quantum Hall effect (FQHE) has been predicted for a two-dimensional (2D) condensate rotating at a frequency matching that of the confining harmonic trap~\cite{Wilkin:1998,Cooper:1999, Wilkin:2000, Paredes:2001,Fischer:2004,Bhongale:2004}. However, the strongly correlated FQHE regime has eluded experimentalists in cold quantum gases due to two problems: it is difficult to confine condensates at rotation speeds matching the trapping frequency and vortex shears destroy condensates at high rotation. A potential solution is the use of a 2D lattice. Introducing a co-rotating optical lattice in the tight-binding regime, in which particles on a lattice site can only tunnel to adjacent sites, provides strong confinement and enhances interactions to enable entry into the strongly-correlated regime. Similar systems have been experimentally demonstrated outside the tight binding regime~\cite{Tung:2006}. In this paper, we present a direct mapping between the angular velocity of a rotating condensate $\Omega$ and the magnetic field as characterized by $\alpha$ in standard condensed matter literature on the quantum Hall effect (e.g.~\cite{Ezawa}). Connections have previously been made to the FQHE for cold atoms in a lattice in the presence of an effective magnetic field~\cite{Jaksch:2003, Palmer:2006} or induced tunneling loops~\cite{Soerenson:2005}. Further, in a recent paper, Umucalilar {\it et.~al.} ~\cite{Umucalilar:2007} presented the phase diagram for bosons in an optical lattice in the presence of an effective magnetic field. 

We study a BEC in a rotating 2D optical lattice using a Bose-Hubbard Hamiltonian modified by the rotation. Observables are computed using exact diagonalization with box boundary conditions. There are two advantages to using box boundary conditions over periodic boundary conditions. First, periodic boundary conditions are suitable only for values of $\alpha$ in a narrow region around rational values~\cite{Palmer:2006} while box boundary conditions can be used to study the system response for both rational and irrational values of $\alpha$. Second, non-periodic elements (such as a trapping potential or a lattice tilt) can be introduced easily. In the case of box boundary conditions, these advantages come with the twin costs of non-negligible boundary effects and limited access as the study of many-particle systems via the exact-diagonalization method quickly becomes intractable with either an increase in lattice size or an increase in the number of particles. 

We use the Kubo formalism to describe the system's current and density responses to a perturbative potential gradient. Using a high-frequency perturbation to overcome finite-size effects, we observe FQHE features in a single particle system. In particular, the system demonstrates plateaus in the transverse resistivity concurrent with dips in the diagonal resistivity for fractional values of $\alpha$. At these same values, in concordance with Jaksch {\it et.~al.}~\cite{Jaksch:2003}, the site number density is modulated with a periodicity $1/\alpha$. Numerical results are also presented for small many-particle systems. In direct analogy with the classical Hall effect, a pileup of particles due to the Coriolis force is seen along the transverse direction. 

The need for theoretical methods to study the strongly-correlated FQHE regime for bosons in a rotating optical lattice is urgent, as experimental capabilities to realize such systems are rapidly coming to bear. Two main characteristics of this regime are site number densities of order unity due to strong interactions and filling factors (particles per vortex)  of order unity needed for the creation of composite particles necessary to observe quantum Hall phenomena. Accordingly, the first experimental requirement is for an optical lattice in the tight binding regime, traditionally with lattice spacing $d \sim 0.5 \mu m$. The second requirement is for the energy associated with rotation to be of the order of the lattice recoil energy or, equivalently, for the associated Larmor radius to be of the same order as the lattice spacing.  Both requirements have been separately satisfied~\cite{Greiner:2002,Schweikhard:2004}. An immediate benefit of such experiments is the measurement of the equivalent of the flux quantum ($e^2/\hbar$ ) constant for mass transport.

This paper is structured as follows: Section II presents a derivation of the modified Bose-Hubbard Hamiltonian used to study the lattice system along with a discussion of current operators. This Hamiltonian is closely connected to that traditionally used to study Bloch electrons in a magnetic field, but a detailed derivation is useful given the new context of cold gases. Section III  describes the Kubo formalism used to study the linear response of the system. Section IV presents single-particle results for large lattices. Section V contains response characteristics for small many-particle systems. The last section discusses experimental implications and the future outlook for this problem. 

\section{Bose-Hubbard Hamiltonian in rotating frame coordinates \label{Section:Hamiltonian}}

The derivation of a modified Bose-Hubbard Hamiltonian using the symmetric gauge is presented in this section. The angular velocity $\Omega$ is mapped onto the parameter $\alpha$ used commonly in quantum Hall literature. The comparison of this system with that of Bloch electrons in a magnetic field is explored by looking at the single-particle energy spectrum. 

\subsection{Derivation}

The system to be described is a cloud of a fixed number of bosons rotating with an angular velocity $\Omega$ about the $z$-axis. This cloud is trapped in a 2D optical lattice co-rotating with the same angular velocity in the presence of an additional, superimposed two-dimensional harmonic trapping potential of frequency $\omega$. For a non-rotating system ($\Omega=0$), the Hamiltonian $\hat{H}_0$ has components corresponding  to the kinetic energy, the lattice and harmonic trapping potentials, and the energy due to interaction between particles. The effect of rotation is included by using time-independent rotating-frame coordinates by means of the transformation, $\hat{H}=\hat{H}_0-\int d{\bf x} \hat{\Phi}^{\dagger}\Omega L_z\hat{\Phi} $~\cite{Landau:Mechanics} where  $\hat{\Phi}$ is a bosonic annihilation field operator describing the atoms and $L_z$ is the angular momentum operator.  The Hamiltonian can then be written in rotating frame coordinates as,
\begin{eqnarray}
  \hat{H}=\int  &d {\bf x}& \hat{\Phi}^{\dagger}\left( -\frac{\hbar ^2}{2m} 
    \nabla ^2  \right. \nonumber\\
    &&\left.  + V ^{\mathrm{lat}} ({\bf x}) + V ^{\mathrm{t}}  ({\bf x})+\frac{g}{2} \hat{\Phi}^{\dagger} \hat{\Phi} -\Omega L_z \right) \hat{\Phi}\,, \label{H1}
\end{eqnarray} 
where $m$ is the mass of a single particle and $g$ is the coupling constant for repulsive two-body scattering via a contact interaction. In this paper, we use a square lattice potential described by $ V ^{\mathrm{lat}}({\bf x})=V _0 (\sin ^2 (\pi x/d)+\sin ^2 (\pi y/d))$. Finally, the trapping potential is $ V ^{\mathrm{t}} ({\bf x})=m\omega^2r^2/2$ with $r\equiv|{\bf x}|$. Equation~(\ref{H1}) can be rewritten as, 
\begin{eqnarray}
\hat{H}=\int &d{\bf x}&\hat{\Phi}^\dagger \left( \frac{{\bf \Pi}^2}{2m}+V ^{\mathrm{lat}} ({\bf x})\right. \nonumber \\ 
&& \left.+\frac{1}{2}m(\omega^2-\Omega^2)r^2 +\frac{g}{2} \hat{\Phi}^{\dagger} \hat{\Phi} \right) \hat{\Phi}\,. \label{H}
\end{eqnarray}
Here, ${\bf \Pi}\equiv -i \hbar \nabla + m{\bf A}({\bf x})$ is the covariant momentum, and  ${\bf A}({\bf x}) \equiv {\bf \Omega}\times {\bf x}$ is the equivalent of a magnetic vector potential stemming from the rotation. 

The field operator $\hat{\Phi}$ can be expanded in several ways. One common expansion for the stationary lattice problem uses Wannier orbitals $W^l_S({\bf x}-{\bf x_i})$, where the sites are indexed by $i$ and the bands by $l$~\cite{Wannier:1962}. If the energy separation between the lowest Bloch band and the first excited band is large compared to the interaction energy and the angular velocity is low ($\hbar \Omega\sim 0.01 E_R$), then a good description is obtained by retaining only Wannier orbitals constructed from the lowest Bloch band, i.~e., $l=0$. With this approximation, the phase description of the single-particle wavefunction is flat within a particular lattice site with sharp gradients at site boundaries. However, for larger angular velocities ($\hbar \Omega \sim 0.1 E_R$), the $\Omega L_z$-term  mixes in higher bands to a non-negligible extent. The primary effect of this mixing is to modify the phase structure within sites. A modified Wannier basis given by
\begin{equation}
W_R({\bf x}-{\bf x_i})\equiv\exp \left( -i\frac{m}{\hbar}\int ^{{\bf x}} _{{\bf x_i}}{\bf A}({\bf x'})\cdot d{\bf x'}\right) W^0_S({\bf x}-{\bf x_i})\label{WR}
 \end{equation}
captures some of this effect by making the azimuthal phase gradient within a site proportional to $\Omega$. The lower limit in the integral is chosen to coincide with the site center. This choice ensures that at ${\bf x}={\bf x_i}$ the Wannier orbital is real, i.~e.,  $W_R(0)=W_S(0)$. A path of integration needs to be chosen such that the basis set defined by $W_R({\bf x}-{\bf x_i})$ satisfies orthonormality. In addition, calculations for the square-lattice problem are greatly simplified if the choice of path allows for each two-dimensional Wannier orbital to be decoupled into a product of one-dimensional Wannier orbitals. One particular choice of a path that meets both criteria is along straight lines parallel to the lattice axes. Note that this path is not unique and decoupling is not required. A quantitative assessment using imaginary time propagation techniques~\cite{Bhat:2006b} shows that the modified Wannier basis set $W_R({\bf x}-{\bf x_i})$ describes the phase gradient within a site better than the regular Wannier basis $W_S({\bf x}-{\bf x_i})$ and captures the pertinent parts of the Hilbert space for our discussion.  

Using the modified Wannier basis $W_R({\bf x}-{\bf x_i})$, the field operator can now be expressed as
\begin{equation}
\hat{\Phi}({\bf x})=\sum _i \hat{a}_i W_R({\bf x}-{\bf x_i})\,, \label{Phi}
\end{equation}
where $\hat{a}_i$ is a site-specific annihilation operator. A modified Bose-Hubbard Hamiltonian is obtained by substituting Eq.~(\ref{Phi}) into Eq.~(\ref{H}), 
\begin{eqnarray}
\hat{H}\! &=&\! -\sum _{\langle i,j \rangle} \left( t+\frac{m(\Omega^2-\omega^2 )}{2}A_1\right) ( \hat{a}^{\dagger}_i \hat{a}_j e^{-i\phi_{ij}}+\hat{a}_i \hat{a}_j ^{\dagger}e^{i\phi_{ij}})  \nonumber \\
&+&\sum_i \left( \epsilon -\frac{m(\Omega^2-\omega^2) }{2}(r_i^2+ A_2) \right) \hat{n}_i  \nonumber \\
&+&\frac{U}{2}\hat{n}_i (\hat{n}_i-1)\,, \label{MBH}
\end{eqnarray}
where $\langle i,j \rangle$ indicates the sum over nearest-neighbor site pairs and $\hat{n}_i\equiv \hat{a}^{\dagger}_i\hat{a}_i$ is the number operator. The definitions for $\phi_{ij},t,\epsilon, A_1$ and $A_2$ follow. The phase for the hopping term is  
\begin{equation}
\phi_{ij}\equiv\frac{m}{\hbar}\int ^{{\bf x_i}} _{{\bf x_j}}{\bf A}({\bf x'})\cdot d{\bf x'}=\frac{m\Omega }{\hbar}(x_iy_j-x_jy_i)\,. \label{phiij}
\end{equation}
Here, the angular velocity $\Omega$ is expressed in units of  $E_R/\hbar$ where $E_R$ is the recoil energy associated with lattice spacing $d$. The parameters $t$ and $\epsilon$ are identical to the hopping and onsite zero-point energies associated with the standard Bose-Hubbard model~\cite{Fisher:1989} and are obtained by evaluating the integrals 
\begin{eqnarray}
  t \! & \equiv &\! \int d{\bf x} W ^{*} _S ({\bf x}-{\bf x _i}) \left( -\frac{\hbar ^2}{2m} \nabla ^2 + V ^{(lat)} ({\bf x}) \right) W _S ({\bf x}-{\bf x _j}) \,, \label{t} \nonumber \\
  \\
  \epsilon \! & \equiv & \! \int d{\bf x} W ^{*} _S ({\bf x}-{\bf x _i}) \left( -\frac{\hbar ^2}{2m} \nabla ^2 + V ^{(lat)} ({\bf x}) \right) W _S({\bf x}-{\bf x _i}) \,.\label{eps} \nonumber \\
\end{eqnarray}   
\begin{figure}[t]
\begin{center}
   \includegraphics[width=8.4cm]{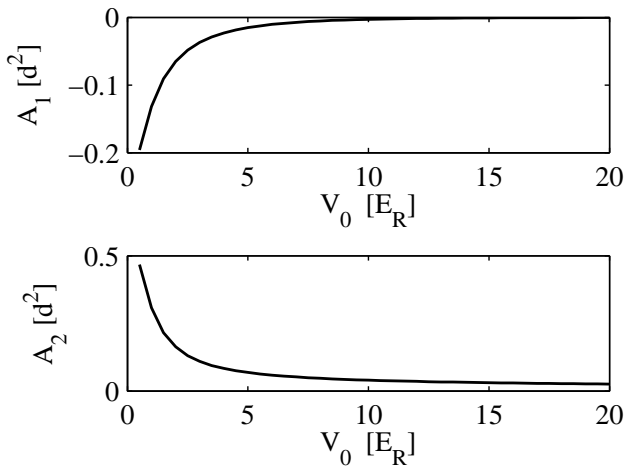}
   \caption{ Overlap integrals $A_1$ and $A_2$ as a function of lattice depth for a standing-wave optical lattice described by  $V^{(lat)}({\bf x})=V _0 (\sin ^2 (\pi x/d)+\sin ^2 (\pi y/d))$. The lattice depth is given in units of the recoil energy $E_R=\hbar^2\pi^2/2md^2$.  $A_1$ and $A_2$ are in units of $d^2$. In the tight-binding regime $\left( V_0/E_R\gtrsim 5 \right)$, $A_1\sim -0.273 \exp \left( -V_0^{0.656}\right)$ and $A_2 \sim 0.368 \exp \left( -V_0^{0.337}\right)$.   \label{A1A2}}
\end{center}
\end{figure}
The modifications to these terms due to rotation are proportional to $(\Omega^2-\omega^2)$ and to  two new overlap parameters given by 
\begin{eqnarray}
A_1&\equiv& \int dx W^*_S(x-x_i) \: \left( x-x_i \right) ^2 \:W_S(x-x_j)\,, \label{A1} \\
A_2&\equiv& 2\int dx W^*_S(x-x_i) \: \left( x-x_i \right) ^2 \: W_S(x-x_i)\,, \label{A2}
\end{eqnarray}
where $W_S(x-x_i)$ is a one-dimensional Wannier function. There is an additional factor of two in $A_2$ because of identical onsite overlaps along the $x$- and $y$- directions. Changes in the lattice potential affect both new parameters (Fig.~\ref{A1A2}). The last term in the Hamiltonian describes the onsite-interaction energy, and for an $s$-wave scattering length $a_s$~\cite{Jaksch:1998, Zwerger:2003},
\begin{eqnarray}
  U\equiv\frac{4\pi a_s \hbar ^2}{m} \int d{\bf x}\left|  W_S({\bf x}-{\bf x_i})\right|^4 \,.\label{U}
\end{eqnarray}   
The interaction term as described in Eq.~(\ref{U}) is included for completeness and is used only implicitly in this paper. The single-particle discussions are trivially independent of $U$, while the many-particle hardcore-boson analyses implement the $U\rightarrow \infty$ condition using the two-state approximation. 

There are two other approximations implicit in our approach. The first is the tight-binding approximation whereby only hopping between adjacent sites is considered. This approximation becomes valid when $V_0$ exceeds $5 E_R$. Our calculations are well in the tight-binding regime with $V_0=10E_R$. The second approximation is the use of infinite-lattice Wannier functions for a finite lattice. Due to this, edge effects are not accounted for correctly though the approximation gets better with increase in lattice size.  

We implement this approach by constructing the Hamiltonian using a truncated Fock-number basis for the desired number of sites and diagonalizing it numerically. Note that we implicitly introduce infinite potential walls around the lattice by spatially restricting particles to a limited number of sites. As will be discussed in Sec.~\ref{Section:Kubo}, we minimize the effect of the box boundary conditions in the linear response by modulating the perturbation with a period much smaller than the time scales associated with particles tunneling from one lattice site to the next.
 
Note that the modified Bose-Hubbard Hamiltonian in Eq.~(\ref{MBH}) is obtained using the symmetric gauge for the vector potential ${\bf A}({\bf x})$. An expression equivalent to that used by Jaksch {\it et.~al.}~\cite{Jaksch:2003} can be obtained using the Landau gauge with ${\bf A}_L ({\bf x})=\Omega y\hat{i}$ and setting the trapping frequency  equal to the angular velocity $(\omega=\Omega)$. The two different Hamiltonians can be connected using the transformation $\hat{H}_L=P_L \hat{H} P_L^{-1}$ where 
\begin{equation}
P_L=\exp \left( i\frac{m\Omega}{\hbar} \int _{{\bf x_0}}^{{\bf x}} x'dy' \right)\,. \label{PL}
\end{equation}
 
\subsection{Mapping angular velocity onto $\alpha$}
\begin{figure}[t]
\begin{center}
   \includegraphics[width=7.0cm]{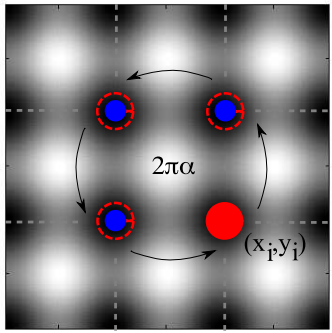}
   \caption{(Color online) Schematic for a particle going around a plaquette in a square lattice. Regions shaded dark correspond to lattice sites and the light regions indicate peaks in the lattice potential. Using Eq.~(\ref{phiij}),  it can be shown that the particle picks up a phase of $2\pi \alpha = 2md^2\Omega/\hbar$ as it returns to its original position as marked by the solid circle. If the path of the particle encloses $P$ plaquettes then the phase picked up is $2\pi \alpha P$. \label{Plaquette}}
\end{center}
\end{figure}

Several two-dimensional problems are characterized by multiply-connected domains where singularities in the topology are typically due to quantized magnetic flux lines (e.~g. Aharonov-Bohm effect~\cite{Aharonov:1959}) or strongly repulsive particles (e.~g. quantum Hall effect~\cite{Klitzing:1980, Tsui:1982}).  In this context, it is useful  to introduce a winding number $2\pi \alpha$ that describes the phase picked up by a particle when it goes around such a singularity. Inaccessible regions in the topology can also be created by means of a suitable potential. 

Consider the lattice potential shown in Fig.~\ref{Plaquette}. The light shaded regions correspond to peaks in the lattice potential that are inaccessible to particles in the tight binding regime. The phase accumulated by a particle adiabatically going around one such simply connected inaccessible region (a plaquette) can be calculated by first breaking the loop into four parts as indicated. For each part, the phase change associated with destroying a particle at a site and creating it in a neighboring site is given by Eq.~\eqref{phiij}. This phase is identical to that associated with the hopping term in the Hamiltonian [Eq.~\eqref{MBH}]. The relationship between the angular velocity $\Omega$ and $\alpha$ is obtained by summing the contributions and is given by
\begin{equation}
\alpha = \frac{md^2}{\pi \hbar} \Omega= \frac{\pi}{2}\left( \frac{\hbar \Omega}{E_R} \right)\,. \label{alpha}
\end{equation}
Henceforth, we will use $\alpha$ to characterize the rotation frequency instead of the angular velocity $\Omega$ in order to maintain the connection with quantum Hall literature. 

\subsection{Single-particle energy spectra}
\begin{figure}[t]
\begin{center}
   \includegraphics[width=7.8cm]{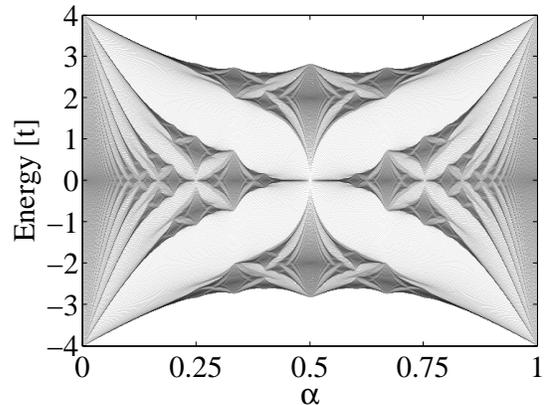}
   \caption{ Single particle energy spectra as a function of the rescaled angular velocity $\alpha$ for a $40\times 40$ lattice. Darker shading indicates greater density of states. The origin of the $y$-axis has been shifted to coincide with the onsite energy.\label{Hofstadter}}
\end{center}
\end{figure}
Having derived the Hamiltonian, a useful crosscheck is the comparison of the single-particle energy spectrum  with that for a Bloch electron in the presence of a magnetic field. The energy spectrum for a single particle in a $40\times 40$ lattice is plotted as a function of $\alpha$ in Fig.~\ref{Hofstadter}. The energy contribution due to the centrifugal force is eliminated by setting the trapping frequency equal to the angular velocity $(\omega=\Omega)$. This is identical to the condition necessary to reach the highly degenerate lowest Landau level (LLL) for the same problem in the absence of a lattice. Applying the LLL condition in the lattice context has two consequences for the energy spectrum: first, the spectrum becomes periodic as a function of $\alpha$ with a periodicity $\Delta\alpha=1$ and second, the spectrum is symmetric about $\alpha=0.5$ and takes on the shape of the Hofstadter butterfly\cite{Hofstadter:1976} --- originally used to describe the energy spectra for an electron in a periodic potential in the presence of a magnetic field. For $\omega \ne \Omega$, both the symmetry of the energy spectra about $\alpha=0.5$ and the periodicity are disrupted as the entire spectrum shifts up or down as a function of $\left( \Omega^2-\omega^2 \right)$. As shown by Analytis {\it et.~al.}~\cite{Analytis:2003}, the fractal nature of the spectra becomes increasingly well-defined as the size of the lattice under consideration grows. 

The grayscale in Fig.~\ref{Hofstadter} describes the density of states and the finite nature of the lattice manifests itself in the sparse energy levels between bands. For $\alpha\ll 1$, the lowest bands are linearly proportional to $\alpha$ leading one to draw comparisons to the Landau energy spectra for a single particle in a 2D harmonic oscillator. The Landau energy levels are given by $E_n=(n+1/2)\hbar \omega _c$ where $n$ is an integer and $\omega _c$ is the cyclotron frequency. For small $\alpha$ in the spectra shown in Fig.~\ref{Hofstadter}, the slopes are not similarly proportional to the band index, e.~g., for the five lowest bands, the slopes are $\sim$ 6, 17, 27, 35 and 43.  

\section{Kubo linear response \label{Section:Kubo}}
The Hall effect describes the longitudinal and transverse transport responses of a two-dimensional electron gas in the presence of a magnetic field to an applied electric potential gradient.  The mapping between magnetic flux density for the 2D electron problem and angular velocity for a rotating gas is valid when the latter problem is formulated in rotating frame coordinates. Accordingly, a potential gradient is introduced in the rotating frame by linearly modifying the lattice onsite energy along the direction of the tilt. This section is divided into two parts. The first part lays out two quantities useful for studying particle transport in this system: the inter-site current operator and onsite density. The second part briefly sketches a derivation of the Kubo formula used to study the linear response of this system. 

\subsection{Current and density operators}
The single particle current in the rotating frame, equivalent to the N\"other current associated with local phase changes of the wavefunction, is obtained by using the mass continuity equation for an infinitesimal volume. The current operator is then realized by quantizing the field using Eq.~(\ref{Phi}) and is given by,
\begin{eqnarray}
\hat{J}^{R}({\bf x})&=&\frac{1}{2m} \left[ 
\hat{\Phi}^{\dagger}({\bf x}) \left(  \frac{\hbar}{i} \nabla -m{\bf A}({\bf x})\right)\hat{\Phi}({\bf x}) \right.\nonumber \\ 
&+&\left. \left( \left(  \frac{\hbar}{i} \nabla -m{\bf A}({\bf x})\right) \hat{\Phi}({\bf x})\right)^{\dagger} \hat{\Phi}({\bf x})
\right]\,. \label{JR}
\end{eqnarray}
The connection to the current operator for the stationary lattice is made by using the transformation  $\hat{\Phi}({\bf x})= P\hat{\Phi}_S({\bf x})P^{-1} $ where,  
\begin{equation}
P\equiv\exp \left( -i\frac{m}{\hbar}\int ^{{\bf x}} _{{\bf x_0}}{\bf A}({\bf x'})\cdot d{\bf x'}\right)\,.\label{P}
\end{equation}
Substituting for  $\hat{\Phi}({\bf x})$ in Eq.~(\ref{JR}) yields, 
\begin{eqnarray}
\hat{J}^{S}({\bf x})&=&\frac{\hbar}{mi} Im \left( \hat{\Phi}_S^{\dagger}({\bf x})\nabla \hat{\Phi}_S({\bf x}) \right) \\
&=& P\left( \hat{J}^{R}({\bf x}) + {\bf A}({\bf x}) \hat{\Phi}^{\dagger}({\bf x})\hat{\Phi}({\bf x})\right) P^{-1}\,, \label{JS}  
\end{eqnarray}
where the second term in Eq.~(\ref{JS}) is needed for conservation of current~\cite{Fetter}. We obtain the current from site $i$ into site $j$ by integrating the N\"other current [Eq.~(\ref{JR})] across the boundary between the two sites, 
\begin{equation}
\hat{J}_{ij}\approx \frac{\hbar\gamma}{im}\left( \hat{a}^{\dagger}_i\hat{a}_j e^{i\phi _{ij}}-\hat{a}^\dagger _j \hat{a}_ie^{-i\phi_{ij}}\right)\,, \label{Jij} \\
\end{equation}
where
\begin{eqnarray}
\gamma &\equiv& \int ^{d/2}_{-d/2} dy W_S^* (y)W_S(y) \nonumber \\
              &&\times \left[ W_S \left( x-\frac{d}{2} \right) \partial _x W_S \left( x+\frac{d}{2}\right)\right] _{x\rightarrow 0}\,. \label{gamma}
\end{eqnarray}
For a lattice depth of $V_0=10E_R$, $\gamma\approx 0.094 d^2$.
 
The onsite density operator for a unit cell $i$ [Fig.~\ref{EndCurrentsSchematic}] is,
\begin{equation}
\hat{\rho}_i\approx B_1 \hat{a}^{\dagger}_i\hat{a}_i+B_2 \sum_{\langle i,j\rangle} \left(  \hat{a}^{\dagger}_i\hat{a}_j e^{\phi_{ij}}+h.c.\right) \label{rho}
\end{equation}
where $B_1$ and $B_2$ are dimensionless overlaps within a unit cell for Wannier functions centered on the same site and adjacent sites respectively. For our calculations well in the tight-binding regime, $V_0=10 E_R$, $B_1\approx 0.9969$ and $B_2\approx 1.6591\times10^{-5}$.    
  
\subsection{Kubo formula}
\begin{figure}
\begin{center}
  \includegraphics[width=7.8cm]{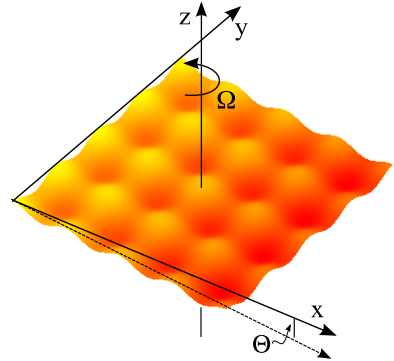}
   \caption{(Color online) Scheme for creating a perturbative linear gradient potential in the system. The lattice potential is tilted by an angle $\Theta$ along the $x$-direction.  The 2D trapping potential is cancelled out by the centrifugal force  at $\omega=\Omega$ and has not been shown. In order to reduce the effect of the implicit infinite boundary potential walls in our calculations, we consider an AC perturbation where the tilt angle is modulated by a frequency $\nu=d\Theta/d\tau$. \label{KuboSchematic}}
\end{center}
\end{figure}

Consider the lattice system sketched in Fig.~\ref{KuboSchematic}. The optical lattice is co-rotating with the condensate about the $z$-axis with angular velocity $\Omega$. The harmonic trap (not shown) frequency is adjusted to $\omega=\Omega$ such that the centrifugal force is cancelled. The perturbation is introduced by tilting the lattice along the $x$-axis in the rotating frame and is modulated by a frequency $\nu$ to induce sloshing. An AC perturbation is switched on at time $\tau=0$. A common mathematical trick to simultaneously extract both quadrature components of the linear response of the system is to use a complex perturbation. In this case, the sine and cosine (phase-shifted) components of $\hat{V}$ will go through and recombine to give a $\exp(i\nu\tau)$ factor in the final result. Accordingly, the perturbation is written as,
\begin{equation}
\hat{V}(\tau)=A \Theta (\tau) e^{i\nu \tau}\hat{X}=A \Theta (\tau) e^{i\nu \tau}\sum _j x_j \hat{n}_j\,, \label{Perturbation}
\end{equation}
where $A$ is the strength of the perturbation, $x_i$ is the $x$-coordinate of site $i$, and $\Theta(\tau)$ is the Heaviside function.  The effect of the implicit infinite boundaries is mitigated by making the time scales associated with the sloshing small compared to that associated with hopping from one site to the next, i.~e., $\hbar \nu >> t$.  A brief sketch of the derivation for the change in the expectation value of an observable $\hat{Y}$ due to the perturbation follows (see Ref.~\cite{Chakraborty, Mahan} for detailed discussions).
 
The density matrix in the interaction picture $\hat{\rho} ^I (\tau)$ can be broken into a time-independent part and the change $\Delta \hat{\rho}^I (\tau)$ due to the perturbation, 
\begin{equation}
\hat{\rho}^I (\tau)= \hat{\rho}_0+\Delta\hat{\rho}^I(\tau)\,, \label{rhoI}
\end{equation}
where the superscript $I$ marks quantities in the interaction picture. The time independent part $\hat{\rho}_0$ corresponds to the density matrix for the unperturbed system. Retaining the first order terms in the Liouville equation of motion for the density matrix provides an expression for the second term in Eq.~\eqref{rhoI}, 
\begin{equation}
\Delta\hat{\rho}^I(\tau)=-\frac{i}{\hbar}\int ^{\tau}_{\infty} e^{-\eta (\tau-\tau')} \left[  \hat{V}^I(\tau'),\hat{\rho}_0\right] d\tau'\,. \label{Deltarho}
\end{equation}
Here $\eta$ is used to fix the boundary conditions and we take the limit $\eta \rightarrow 0+$ at the end of the calculation. The expectation value of $\hat{Y}$ is,
\begin{equation}
\langle \hat{Y}(\tau)\rangle = Tr\left\{ \hat{Y}(\tau)\hat{\rho}(\tau) \right\}=Tr\left\{ \hat{Y}^I(\tau)\hat{\rho}^I(\tau) \right\}\,.
\end{equation}
The expectation value of the response to the perturbation is,
\begin{equation}
\langle \Delta \hat{Y}(\tau)\rangle=Tr\left\{ \hat{Y}^I(\tau)\Delta \hat{\rho}^I(\tau) \right\}\,.
\end{equation}
At low temperatures, the only contribution to the trace comes from the ground state, i.e., $\hat{\rho}_0  \approx \left|\psi_0\rangle \langle \psi _0 \right| $. The final expression for the expectation value of the response is obtained using this approximation and substituting for $\Delta\hat{\rho} ^I (\tau)$ [Eq.~(\ref{Deltarho})],
\begin{align}
\langle \Delta\hat{Y}&(\tau)\rangle =\frac{Ae^{i\nu \tau}}{\hbar}\sum _{n>0}   \nonumber \\ 
&\left[ \langle \psi_0 |\hat{Y} |\psi _n \rangle \langle \psi_n |\hat{X} | \psi _0 \rangle  \right.  
 \frac{e^{i(\omega _n -\omega _o +\nu)\tau-\eta \tau }-1}{(\omega_n-\omega_o+\nu)-i\eta }  \nonumber \\
 &+ \langle \psi_0 |\hat{X} |\psi _n \rangle \langle \psi_n |\hat{Y} | \psi _0 \rangle  
 \left. \frac{e^{-i(\omega _n -\omega _o -\nu)\tau-\eta \tau}-1}{(\omega_n-\omega_o-\nu)-i\eta} \right], \label{Kubo}
\end{align}
where $|\psi _n\rangle$ are energy eigenstates. Note that $n>0$, i.e., $\langle \psi_0\left| \hat{X}\right|\psi_0 \rangle=0$ , because the unperturbed ground state is symmetric about the $y$-axis while $\hat{X}$ is not. For the purposes of this paper, the linear response [Eq.~\eqref{Kubo}] is evaluated at limits $\tau\rightarrow\infty$, $\eta\rightarrow 0+$ and the prefactor $\exp (i\nu\tau)$ is excluded from results shown. 

\section{Single-particle response \label{Section:SingleParticle}}
This section presents numerical results for the linear transport response of a single particle in a rotating $40 \times 40$ lattice. The system is subjected to a perturbation modulated at frequency $\nu=E_R/\hbar$. The linear response is characterized in terms of the change in end currents and the sample averaged resistivity.

\begin{figure}
\begin{center}
   \includegraphics[width=7.0cm]{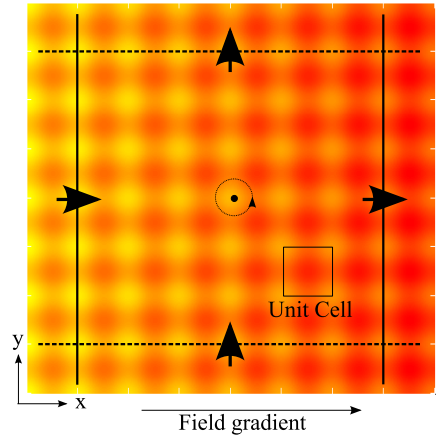}
   \caption{(Color online) Top view schematic depicting end currents for a $8 \times 8$ lattice. The arrows crossing the solid lines mark the longitudinal end currents $\langle \hat{J}^E_x \rangle$ while the arrows crossing the dashed lines indicate the transverse end currents $\langle \hat{J}^E_y\rangle$. The lattice is tilted to the right. \label{EndCurrentsSchematic}}
\end{center}
\end{figure}
 
The ideal way to study the current and voltage characteristics of the system would be to  connect it to reservoirs and compute currents between the system and reservoirs, as is done in studying open quantum systems. However, this approach becomes numerically intractable for systems of size $> 4-6$ sites~\cite{Pepino:2007}. For an isolated system, proxies for the in and out current response of the system are the end currents --- the current response of the system very close to the boundaries of the system. Note that this is true only for a linear response study. The operators for the end currents are obtained by summing current operators across end links as shown in Fig.~\ref{EndCurrentsSchematic}. End currents along each direction are added on either side of the lattice in order to capture only additional currents due to the perturbation. The underlying currents/circulation of the system due to the rotation have been described elsewhere~\cite{Bhat:2006a}. 

\begin{figure}
\begin{center}
  \includegraphics[width=9.0cm]{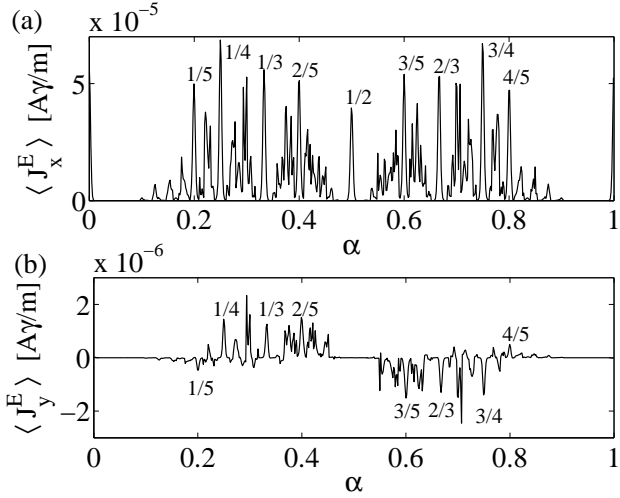}
  \caption{Expectation value of end currents, (a) $\langle \hat{J}^E_x \rangle$ (along the $x$-direction) and (b) $\langle \hat{J}^E_y \rangle$ (along the $y$ direction), as a function of $\alpha$ for a single particle in a $40 \times 40$ lattice subject to a linear ramp perturbation of amplitude $A$ modulated with a frequency $\nu=E_R/\hbar$.\label{EndCurrents}}
\end{center}
\end{figure}

The expectation values for the end currents along the longitudinal and transverse directions are plotted as a function of the winding rate $\alpha$ [Eq.~\eqref{alpha}] in Fig.~\ref{EndCurrents}. The longitudinal end current displays well-defined peaks at fractional values of $\alpha$. At these fractional values of $\alpha=p/q$ (where $p,q$ are integers), the energy spectra [Fig.~\ref{Hofstadter}] breaks up into exactly $q$ bands~\cite{Hofstadter:1976}. For low perturbation frequencies ($\hbar \nu\sim  t$), the denominator in Eq.~\eqref{Kubo} is very small for nearly degenerate states within the same band and the linear response is, in general, large. The system described in this paper has implicit infinite potential walls and the perturbation frequency is far off resonance ($\hbar \nu=E_R\sim 50t$) in order to eliminate Bloch oscillations. Therefore, the denominator in Eq.~\eqref{Kubo} does not become resonant for any value of $\alpha$. The peaks appear due to bigger off-diagonal current matrix elements at fractional values of $\alpha$ and are small due to their non-resonant character. For high frequencies, the height of the peaks goes as $1/\nu$. 

The plot of the transverse current [Fig.~\ref{EndCurrents}(b)] displays peaks/dips at the same values of $\alpha$. The transverse end current is antisymmetric about $\alpha=0.5$. To understand this, consider a value of $\alpha=1-\beta$. The corresponding angular velocity is $\Omega=(\pi\hbar/ Md^2) (1-\beta)$ [Eq.~(\ref{alpha})]. The phase picked up by a particle going around a plaquette is $2\pi \alpha = (2\pi-2\pi\beta) \equiv -2\pi\beta$ [Fig.~\ref{Plaquette}]. The latter phase winding can equivalently be created by rotation in the opposite direction with angular velocity $\Omega=-(\pi\hbar/ Md^2)\beta$ for which the Coriolis force ($\sim {\bf v}\times {\bf \Omega}$) is in the opposite direction.

\begin{figure}
\begin{center}
  \includegraphics[width=8.0cm]{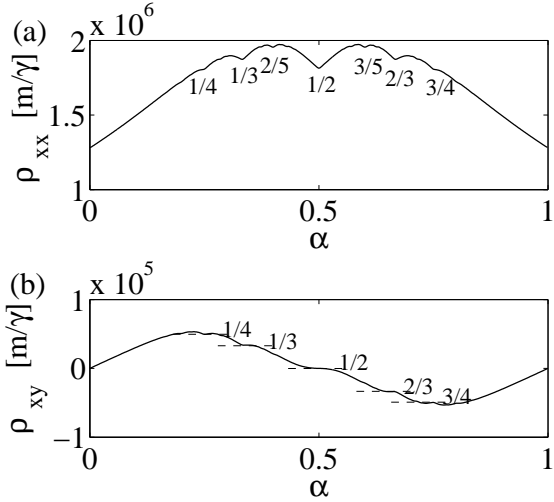}
   \caption{(a) Diagonal and (b) transverse resistivity [Eq.~(\ref{rho})] as a function of angular velocity  for a single particle in a $40 \times 40$ lattice subject to a linear-ramp perturbation of amplitude $E$ modulated with a frequency $\nu=E_R/\hbar$. The shape of the diagonal resistivity plot resembles the cross-section of the Mott Insulator lobe at  $\mu/U=0.5$ as seen in Ref.~\cite{Umucalilar:2007}.   \label{Resistivity}}
\end{center}
\end{figure}

As the size of the lattice under consideration gets bigger, the peak structure in Fig.~\ref{EndCurrents} becomes more well-defined in two ways. First, the peaks become narrower as they get centered closer to exact fractional values of $\alpha$, and second, more peaks appear at other fractional values of $\alpha$. Both of these effects correspond to better resolution of the fractal nature of the energy spectra with bigger lattice size.  The height of the peaks, however, decreases exponentially with lattice size. For example, consider the current plotted in Fig.~\ref{EndCurrents}. The height of the central peak goes as $\sim 1.7\exp (-0.6 L)$ where  $L$ is the number of sites along a side of the lattice.

\begin{widetext}
\begin{figure}
\begin{center}
   \begin{minipage}{18.0 cm}
         \includegraphics[width=4.4cm]{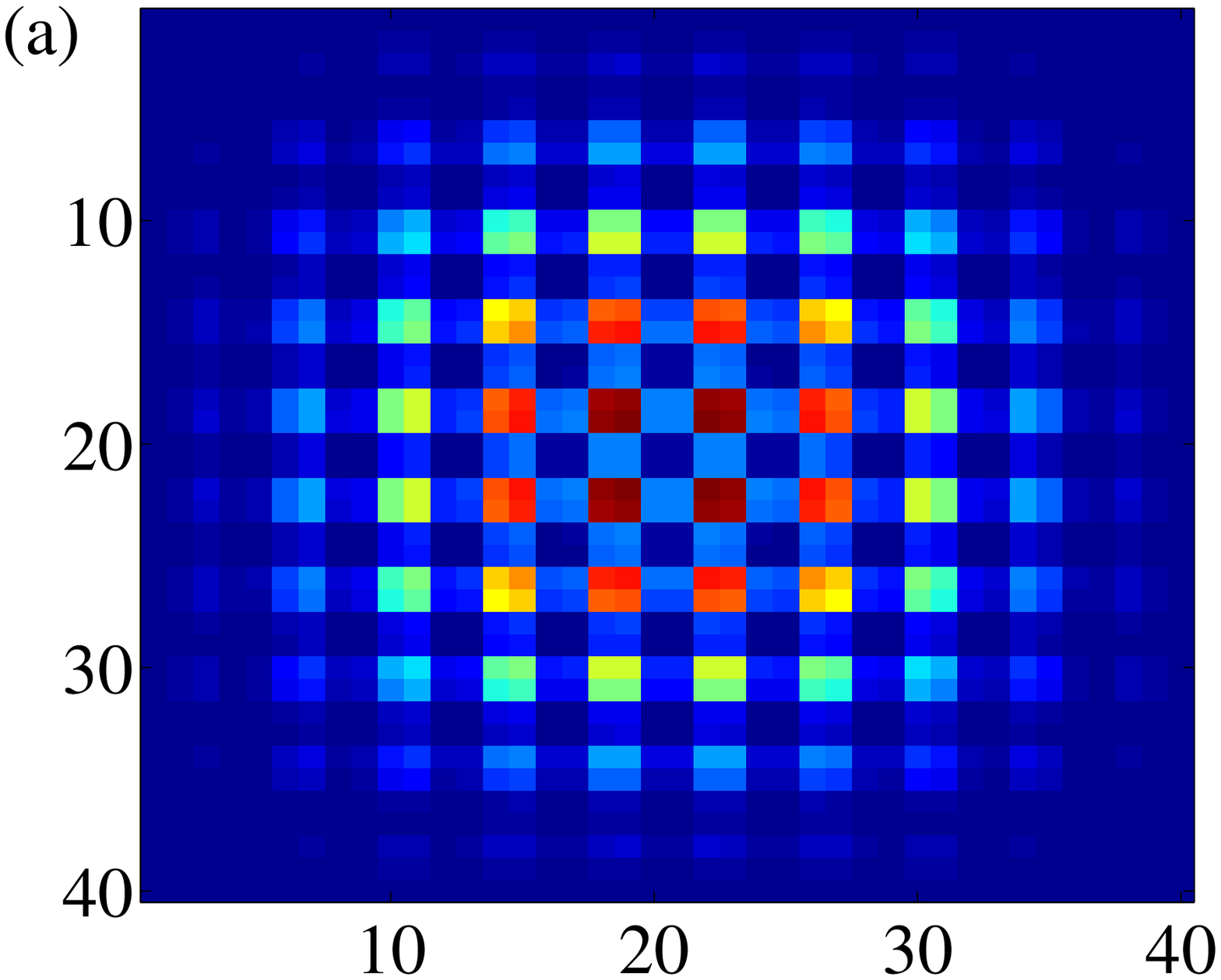} 
	 \includegraphics[width=4.4cm]{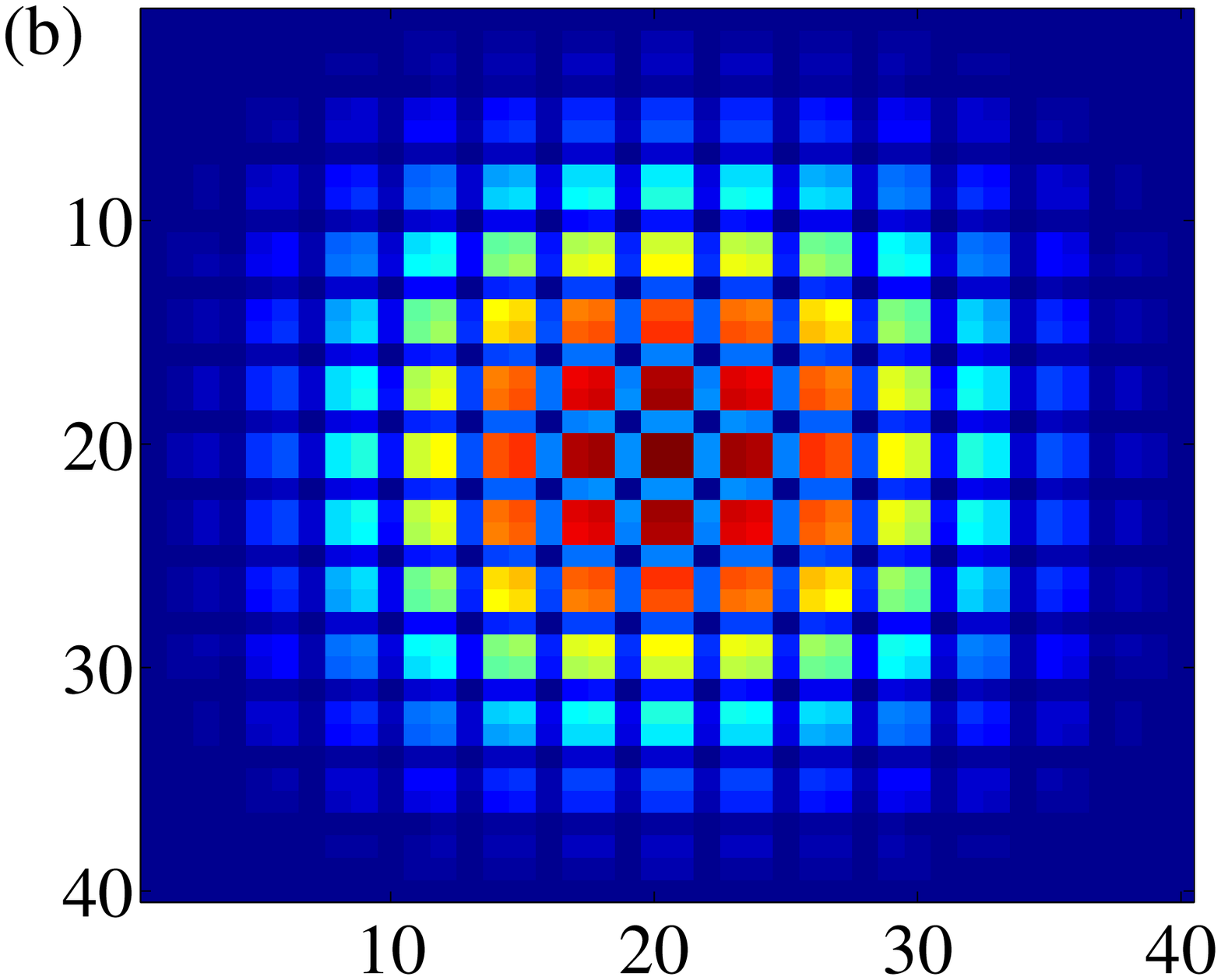}
	 \includegraphics[width=4.4cm]{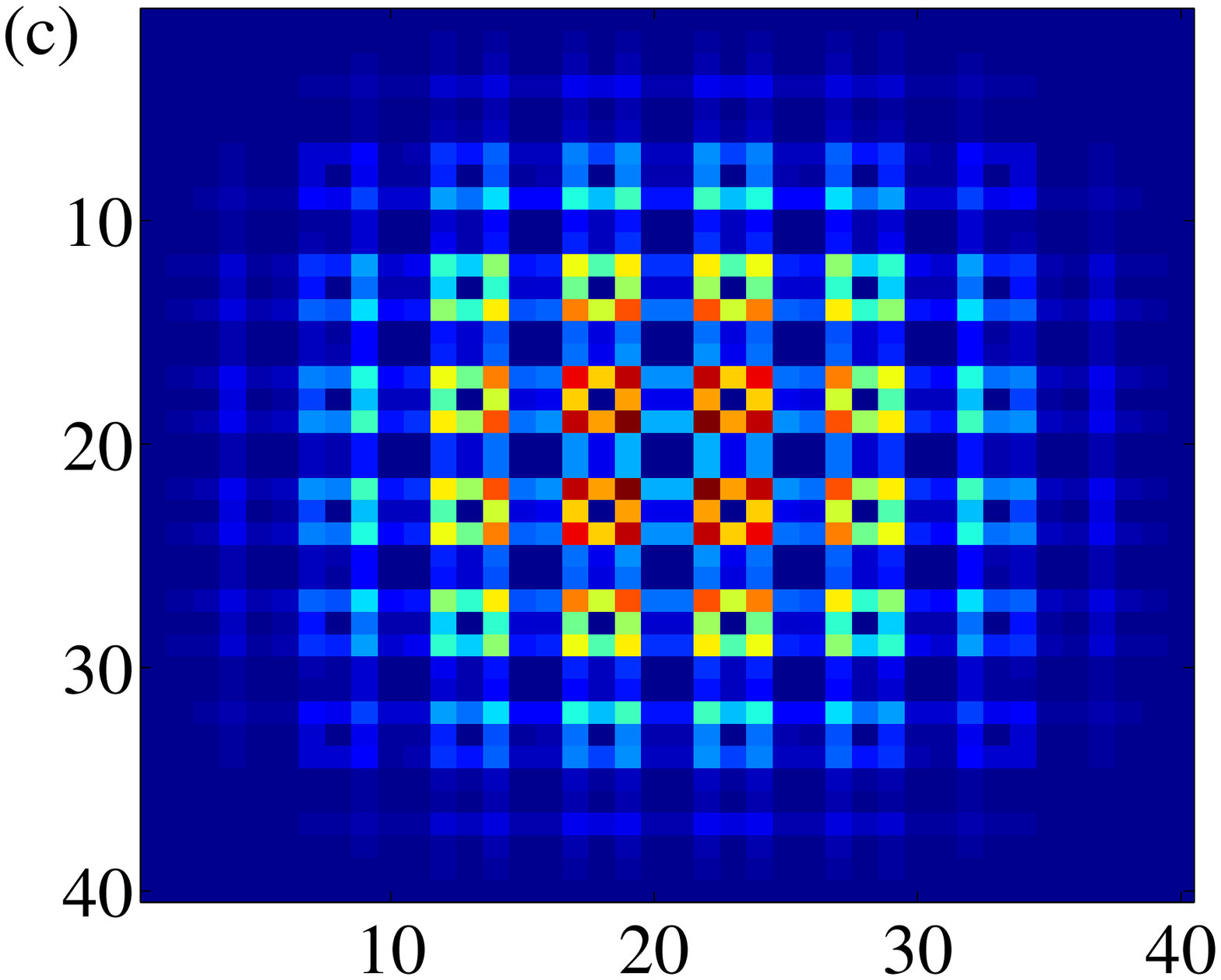}
	 \includegraphics[width=4.4cm]{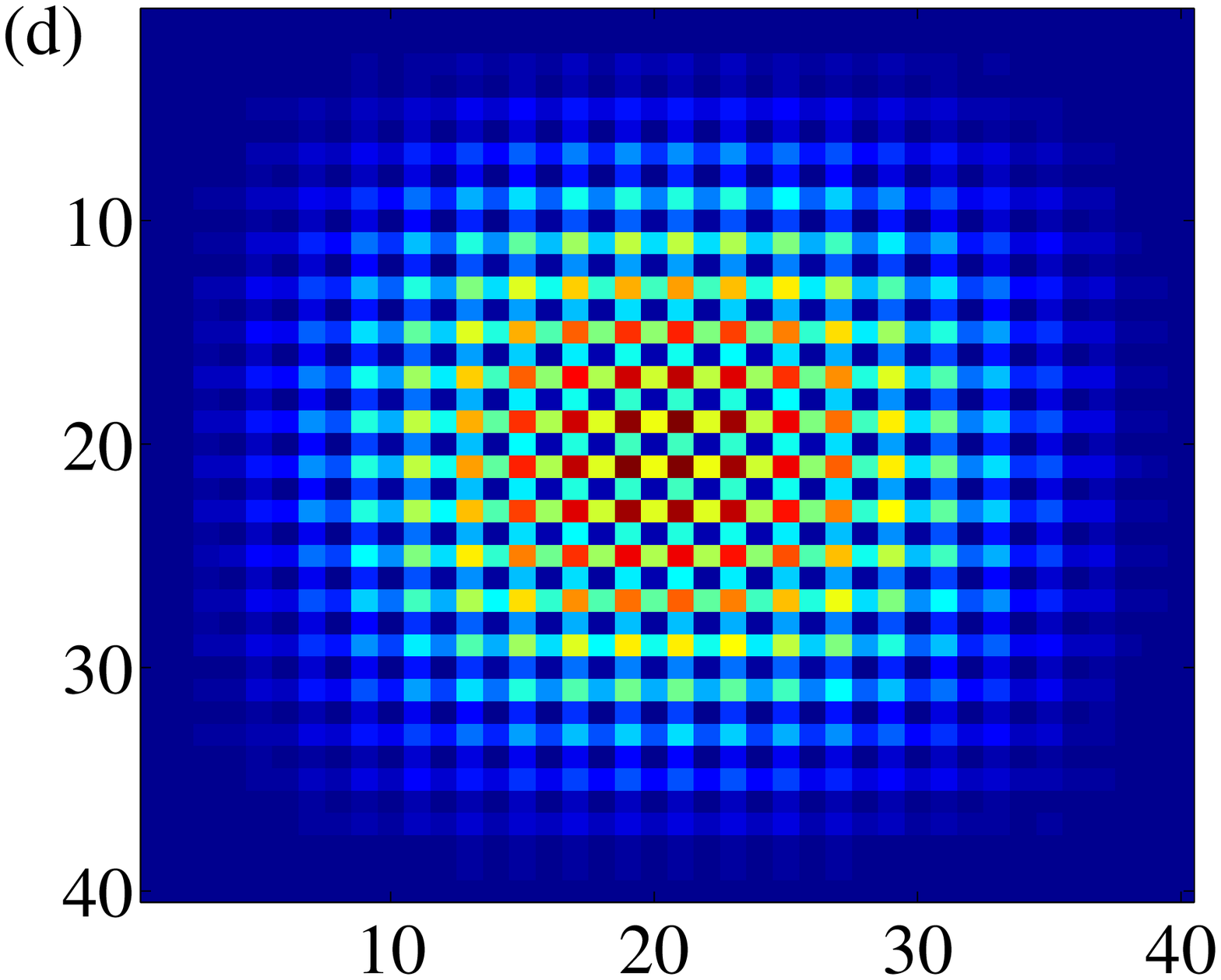}
  \end{minipage}
  \begin{minipage}{18.0 cm}
         \includegraphics[width=4.4cm]{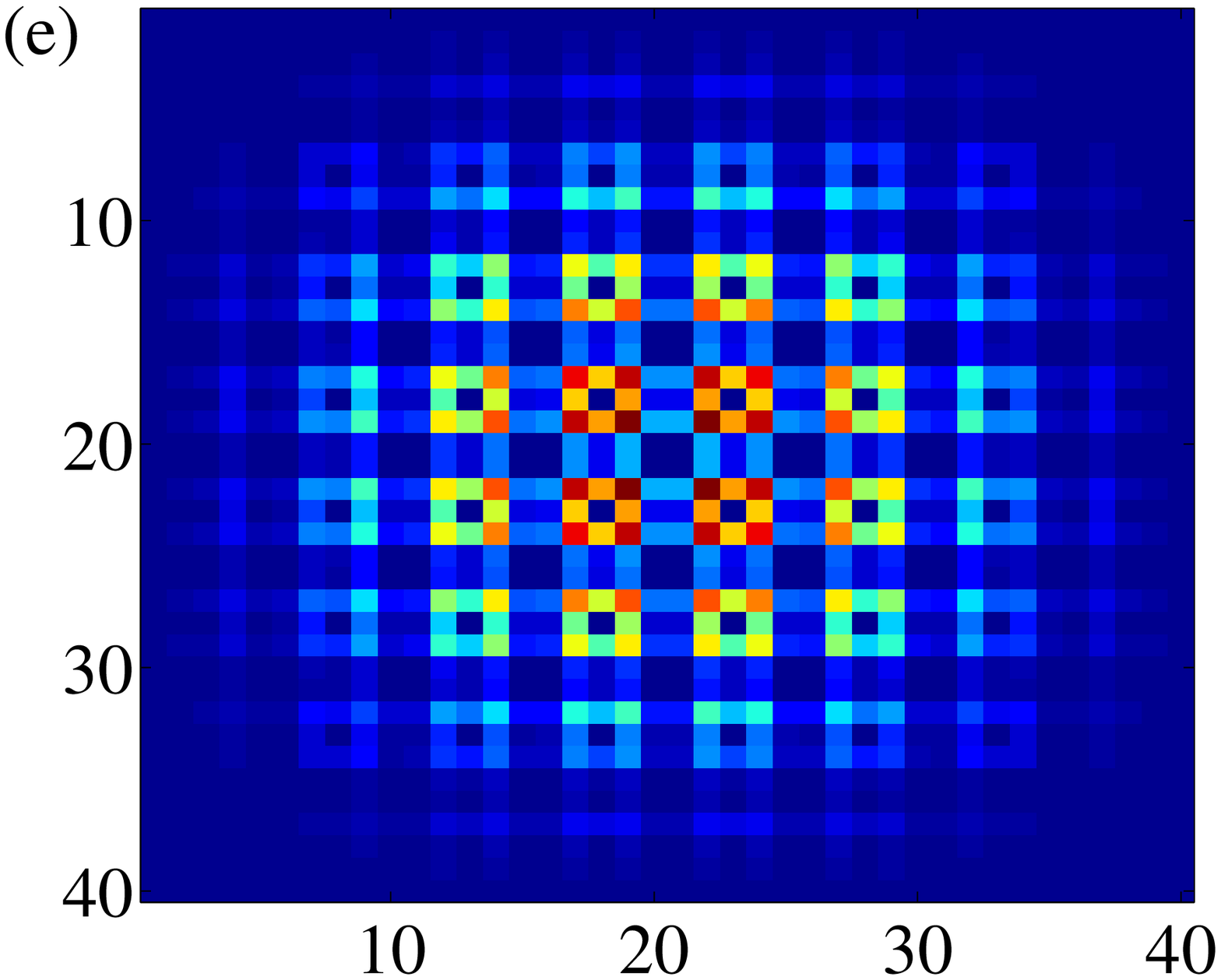} 
	 \includegraphics[width=4.4cm]{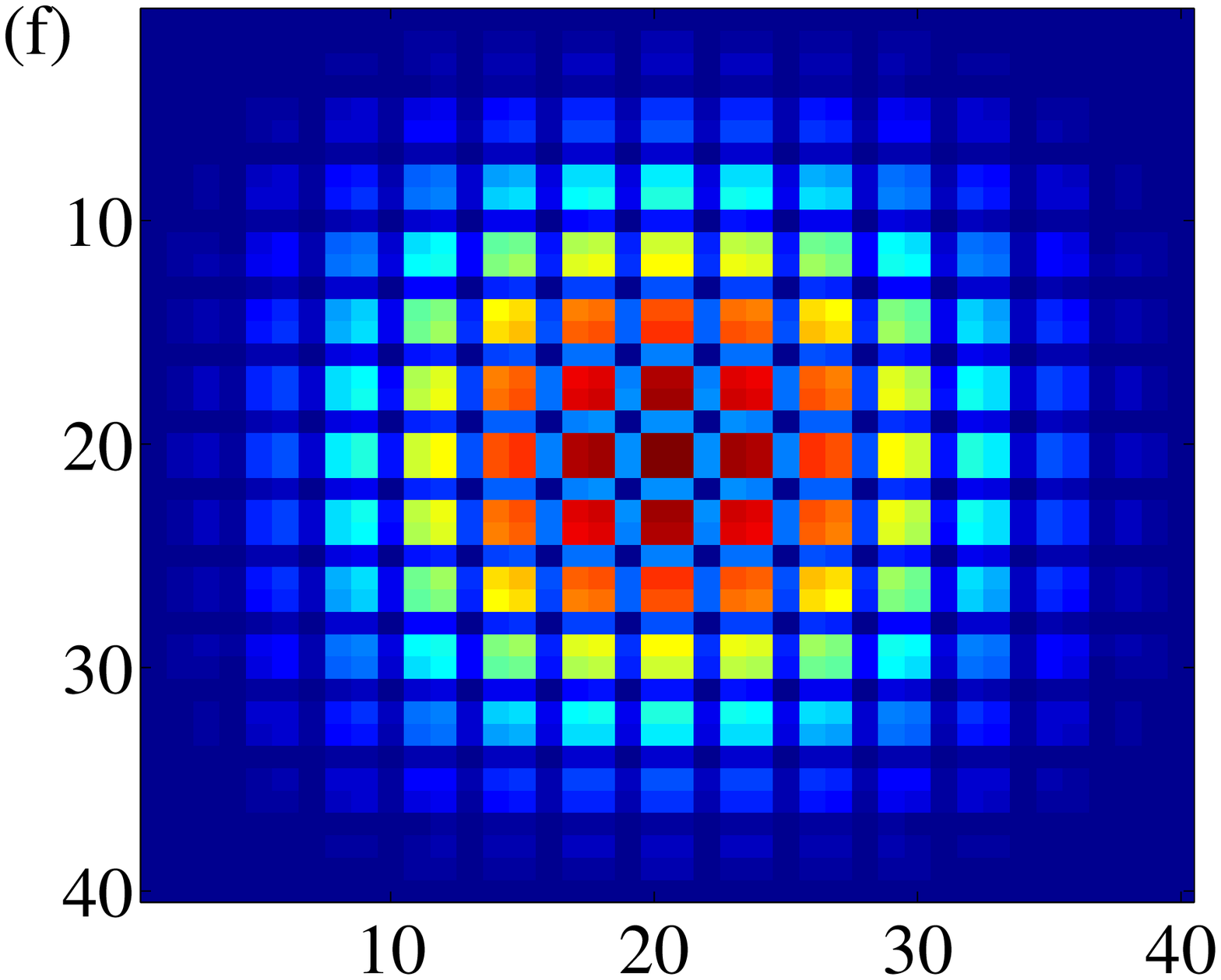}
	 \includegraphics[width=4.4cm]{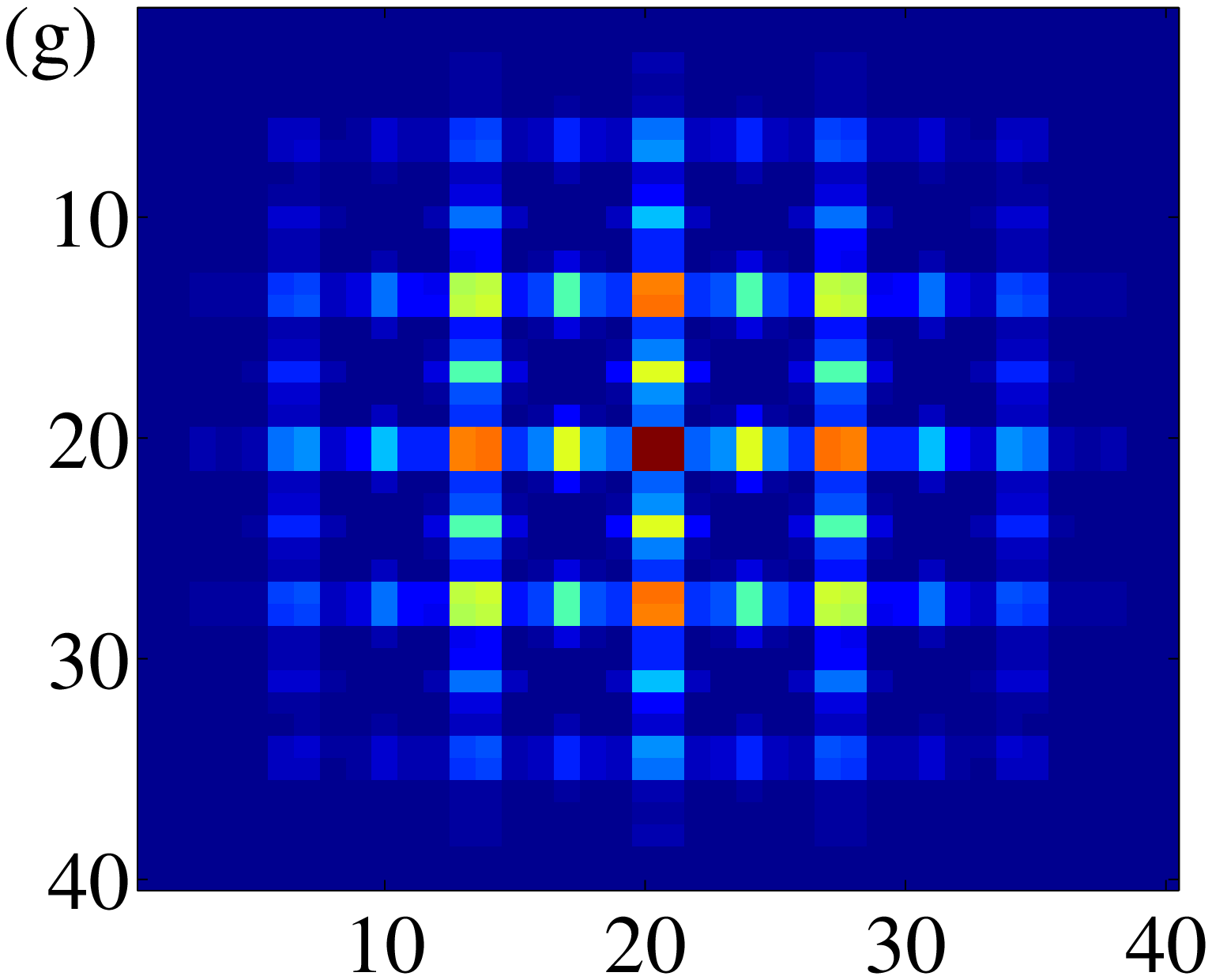}
	 \includegraphics[width=4.4cm]{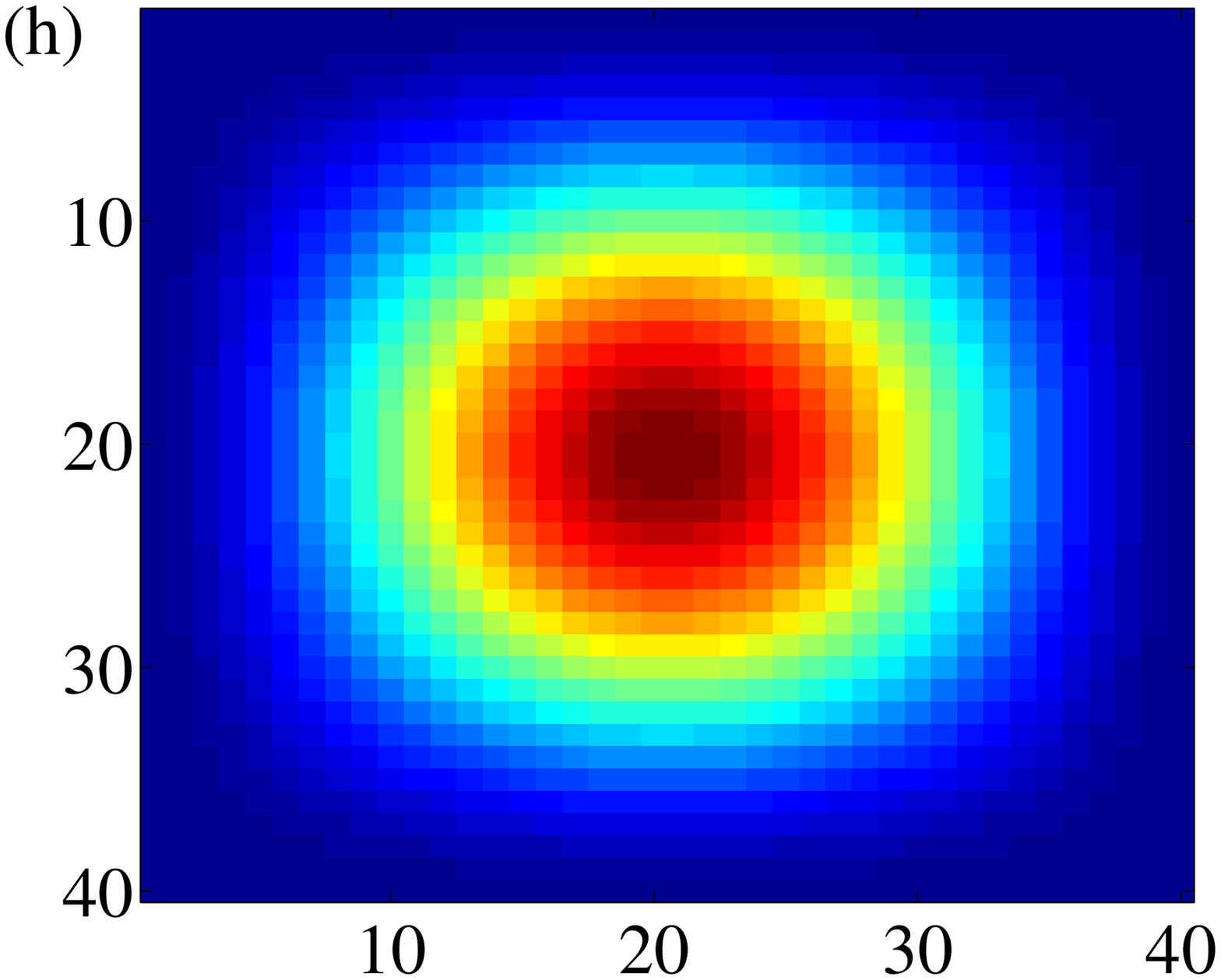}
 \caption{Unperturbed ground state site number-density distributions for a single particle in a $40\times40$ lattice for (a) $\alpha=1/4$, (b) $\alpha=1/3$, (c) $\alpha =2/5$, (d) $\alpha =1/2$, (e) $\alpha = 3/5$, (f) $\alpha =2/3$, (g) $\alpha=\pi/11$, and (h) $\alpha=1$. 
 For simple fractions such as $\alpha=1/4$ (a) or $\alpha=1/3$ (b), the site number-density distribution has peaks separated by 4 and 3 sites respectively. For a fractional value such as $\alpha=2/5$ (c), the  number distribution has periodically arranged rings with centers separated by 5 sites [Fig.~\ref{SiteNumberDensity}(c)].  The density distributions for any value of $\alpha$ are the same as those for $1-\alpha$ (compare (b) and (f) or (c) and (e)). For non rational values, e.~g.~$\alpha=\pi/11$ (g), the periodicity is complicated, if not destroyed. The 2D Gaussian-like envelope seen in all the subplots are due to 2D box infinite wall conditions. \label{SiteNumberDensity}}
  \end{minipage}
\end{center}
\end{figure}
\end{widetext}

A spatial average of the current response across the system smoothens out the peaks in Fig.~\ref{EndCurrents}. The conductivity tensor describes the response of the sample averaged current. If the perturbation is along the $x$-direction, the conductivity tensor elements are 
\begin{equation}
\sigma _{x\mu}=\frac{\langle \Delta \hat{J}_{\mu}\rangle}{A}\,, \label{sigma}
\end{equation}
where $\Delta \hat{J}_{\mu}$ indicates linear response of the sample averaged current along the $\mu$ direction ({\it i.~e.}, the total response of all current operators for links along the $\mu$ direction). The resistivity tensor elements are derived from the conductivity tensor using,
\begin{equation}
\rho _{x\mu}=\frac{\sigma_{x\mu}}{\sigma_{xx}^2+\sigma_{xy}^2}\,. \label{rho}
\end{equation}
The sample-averaged longitudinal and transverse resistivities are shown as a function of $\alpha$ in Fig.~\ref{Resistivity}. The plot of the longitudinal resistivity has dips at all fractional values of $\alpha$, though the dips are now seen only around prominent fractions such as $\alpha=1/2,1/3,2/3,\ldots $. These fractions are the most common in the sense that for a given range of integers, these fractions can be constructed in the most number of ways. The plot of the transverse resistivity shows plateaus at values of $\alpha$ corresponding to these dips. Both features are signatures of the FQHE seen in a 2D electron gas. This appearance of a many-particle effect in a single-particle system is intriguing. A tentative explanation is given by considering the effect of the optical lattice. In a 2D electron gas, the combined effect of the magnetic field and the Coulombic interaction is to arrange the electrons into a lattice. For a filling factor of one, the electrons fill the lowest Landau level forming a hexagonal lattice in the nearest neighbor approximation. The lattice spacing is $2\sqrt{\pi/3}l_B$, where $l_B\equiv\sqrt{\hbar/eB_{\perp}}$ is the magnetic length determined by a magnetic field $B_{\perp}$~\cite{Ezawa}. In addition, the two-dimensional geometry in which electrons cannot cross each other leads to a change in phase equal to $2\pi\alpha$ each time one electron circles another. These effects are reproduced when a lattice is introduced in such a way that the particle picks up a phase of $2\pi \alpha$ going around a plaquette~[Fig.~\ref{Plaquette}]. In Fig.~\ref{SiteNumberDensity}, square periodic density structures are seen for the single-particle case at certain values of $\alpha$ in where the expectation value of the ground-state site number density for the unperturbed system has been defined as  $\langle \hat{n}_i\rangle =\langle \hat{a}^{\dagger}_i\hat{a}_i\rangle $.  At values of $\alpha=p/q$ ($\{p,q\} \in$ integers) corresponding to dips in the longitudinal resistivity, $\langle \hat{n}_i\rangle $ has a periodicity $q=\pi \hbar/md^2\Omega$. This stems from the periodicity due to the hopping term in the Hamiltonian [Eq.~\eqref{H}] (see also Refs.~\cite{Soerenson:2005,Palmer:2006,Umucalilar:2007}). Bragg scattering is a promising probe for such structures~\cite{Jaksch:2003}. Note that the periodicity goes as $1/\Omega$ and not as $1/\sqrt{\Omega}$ as might be expected by direct comparison with the electron gas system.

As discussed in Ref.~\cite{Palmer:2006}, for $\alpha\ll 1$ in an infinite system, the length scale of the wavefunctions is much larger than the lattice spacing and in this continuum limit, the ground state of the system is the one-half Laughlin state. In addition, the site number density distributions for $\alpha$ and $1-\alpha$ are identical (e. g., Figs.~\ref{SiteNumberDensity}(b) and (f)). Therefore, the number density distribution for $\alpha=1$ is the same as that for $\alpha=0$ and corresponds to a system without rotation or lattice. The concentration of particles at the center [Fig.~\ref{SiteNumberDensity}(h)] is due to the 2D infinite box potential. 

\section{Many-particle response\label{Section:ManyParticles}}
The introduction of more hard-core bosons to the two-dimensional system described earlier adds an additional degree of freedom to the problem. In the single-particle system, circling a plaquette added a phase of $2\pi\alpha$. This is still true in the many-particle system but there is an additional phase when two particles are exchanged~\cite{Khare}, i.~e., $\Psi({\bf x}_1,{\bf x}_2)=\exp (i2\pi \alpha') \Psi({\bf x}_2,{\bf x}_1)$, where $\alpha '$ is a dimensionless winding rate similar to $\alpha$. This substantially complicates the problem. This section extends the earlier analysis to many particle systems using numerical results for small systems. 

\begin{figure}
\begin{center}
 \includegraphics[width=7.8cm]{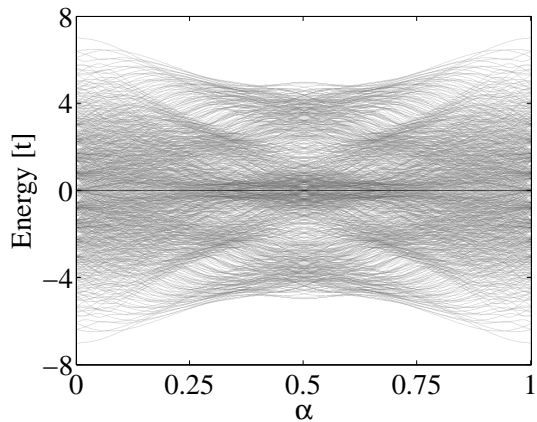}
   \caption{ Energy spectrum for two particles in a $8\times 8$ lattice. The gray shading marks the density of states on a logarithmic scale. The trapping frequency $\omega$ is set equal to the angular velocity $\Omega$. \label{ManyParticleEnergySpectra}}
\end{center}
\end{figure}

The energy spectrum for two hardcore bosons in a $8\times 8$ lattice is plotted in Fig.~\ref{ManyParticleEnergySpectra}. The overall butterfly outline seen in Fig.~\ref{Hofstadter} is preserved. For $N$ particles, the total energy band width defined by the maximum energy difference at $\alpha=0$, is $\Delta E_{max}=8tN$. The gray shading describes the density of states which is marked by degeneracy at energy $E=0$. It is difficult to delineate a band structure due to finite lattice size. However, at the most distinct regions ($\alpha\sim 0.5 $), there appear to be three bands as opposed to two seen in Fig.~\ref{Hofstadter}. Diagonalization of  larger many-particle systems quickly becomes intractable due to the exponential scaling of the Hilbert space dimension with particle number. 

\begin{figure}
\begin{center}
   \begin{minipage}{8.5 cm}
         \includegraphics[width=4.0cm]{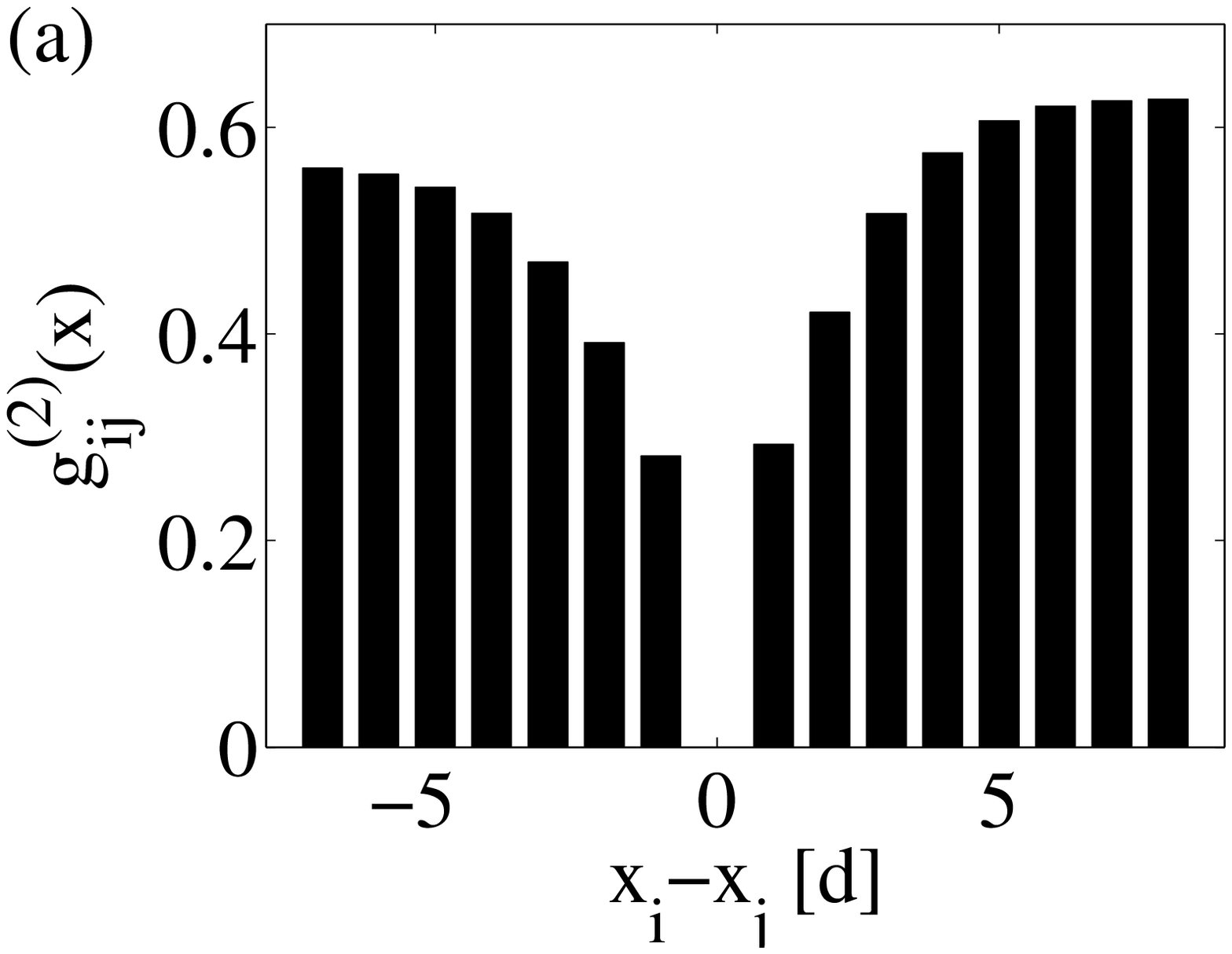} 
	 \includegraphics[width=4.0cm]{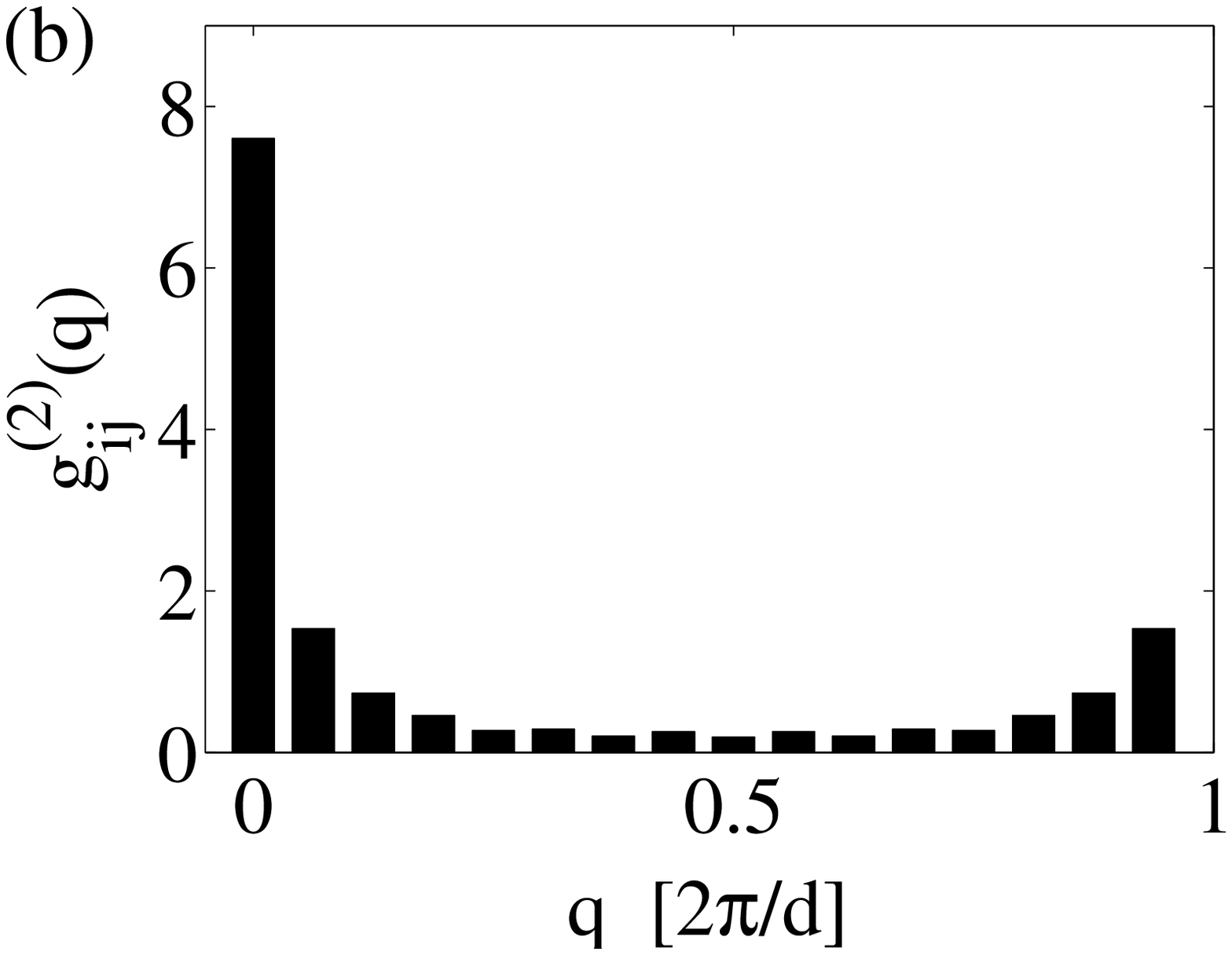}
  \end{minipage}
  \begin{minipage}{8.5 cm}
         \includegraphics[width=4.0cm]{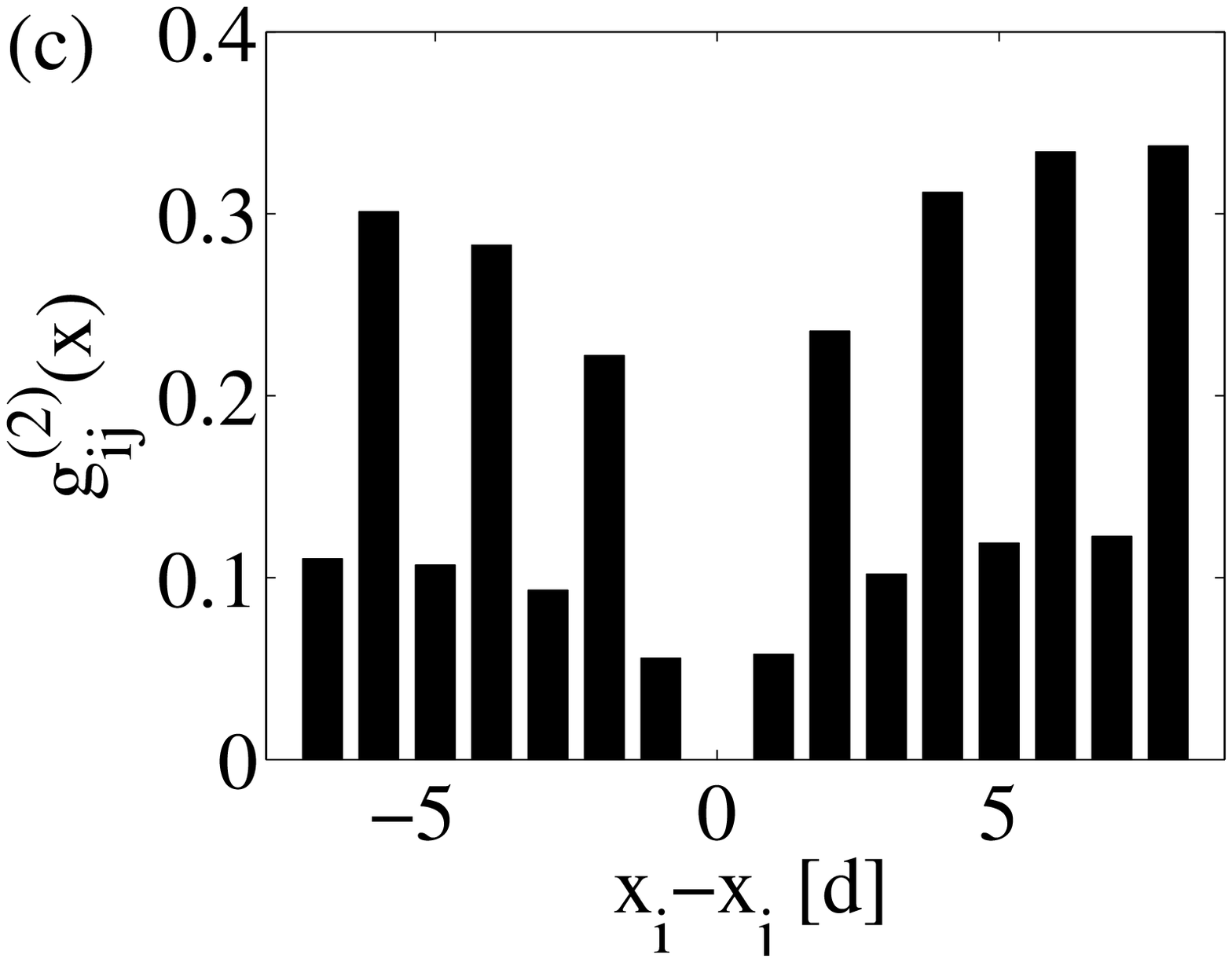} 
	 \includegraphics[width=4.0cm]{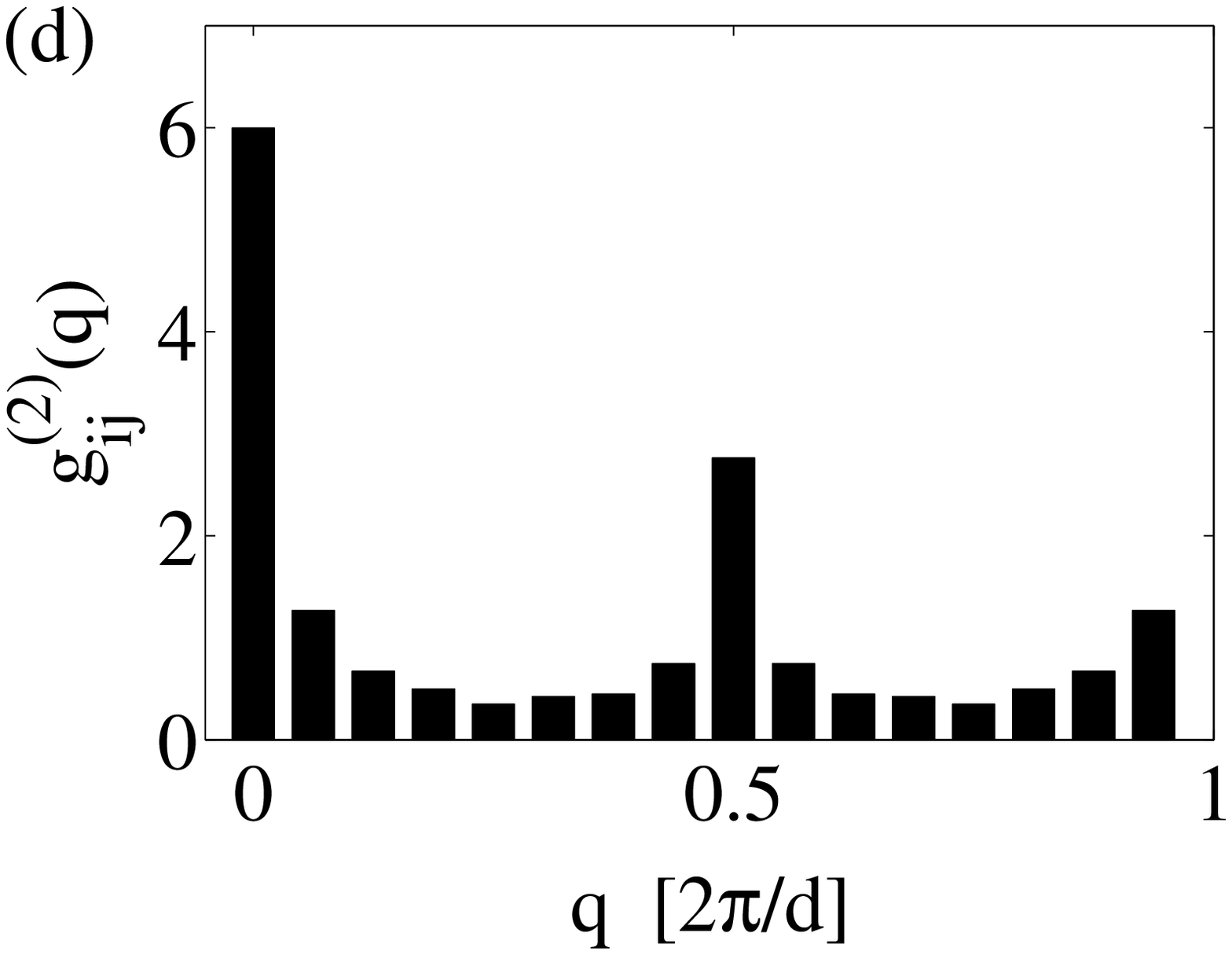}
  \end{minipage}
 \begin{minipage}{8.5 cm}
         \includegraphics[width=4.0cm]{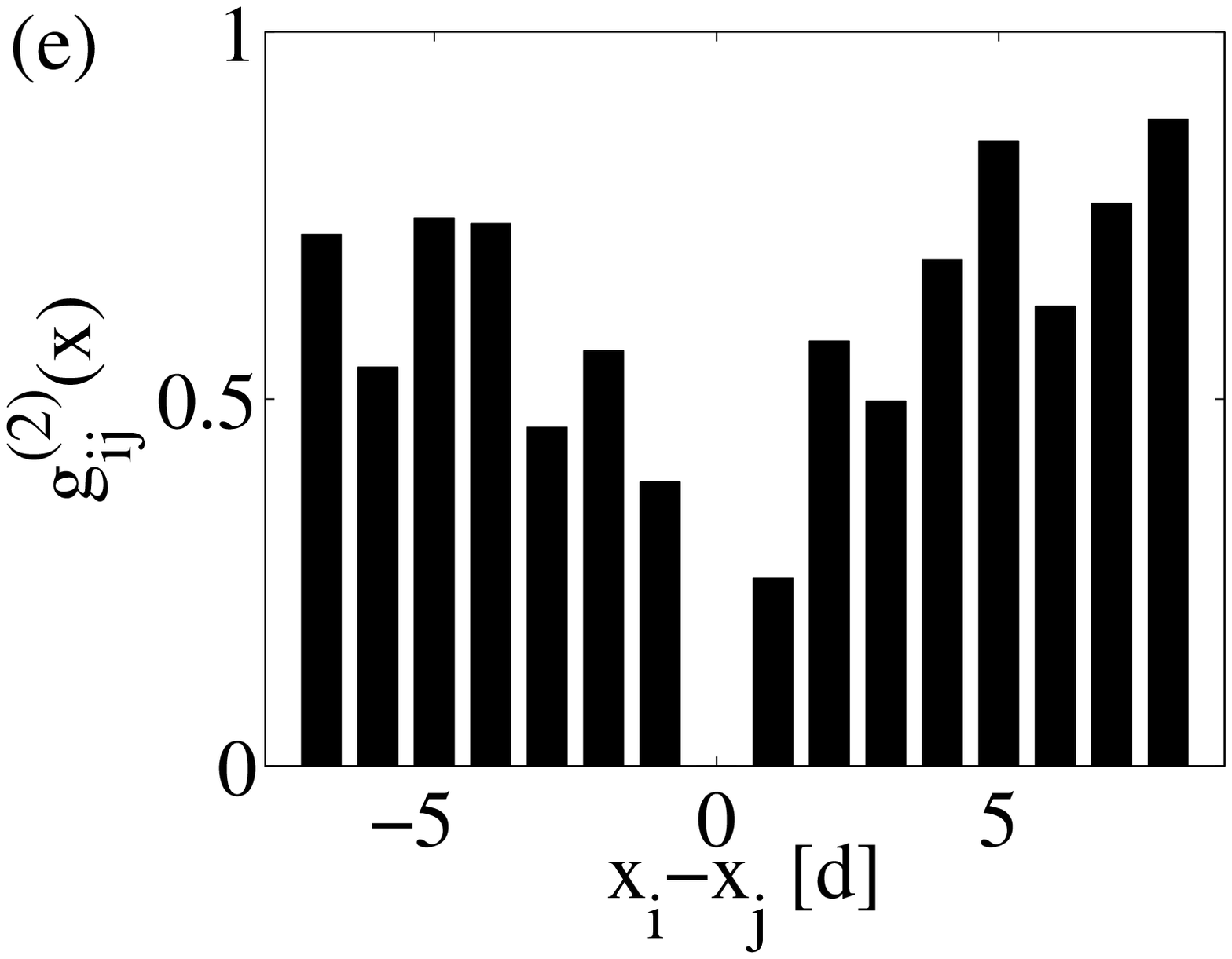} 
	 \includegraphics[width=4.0cm]{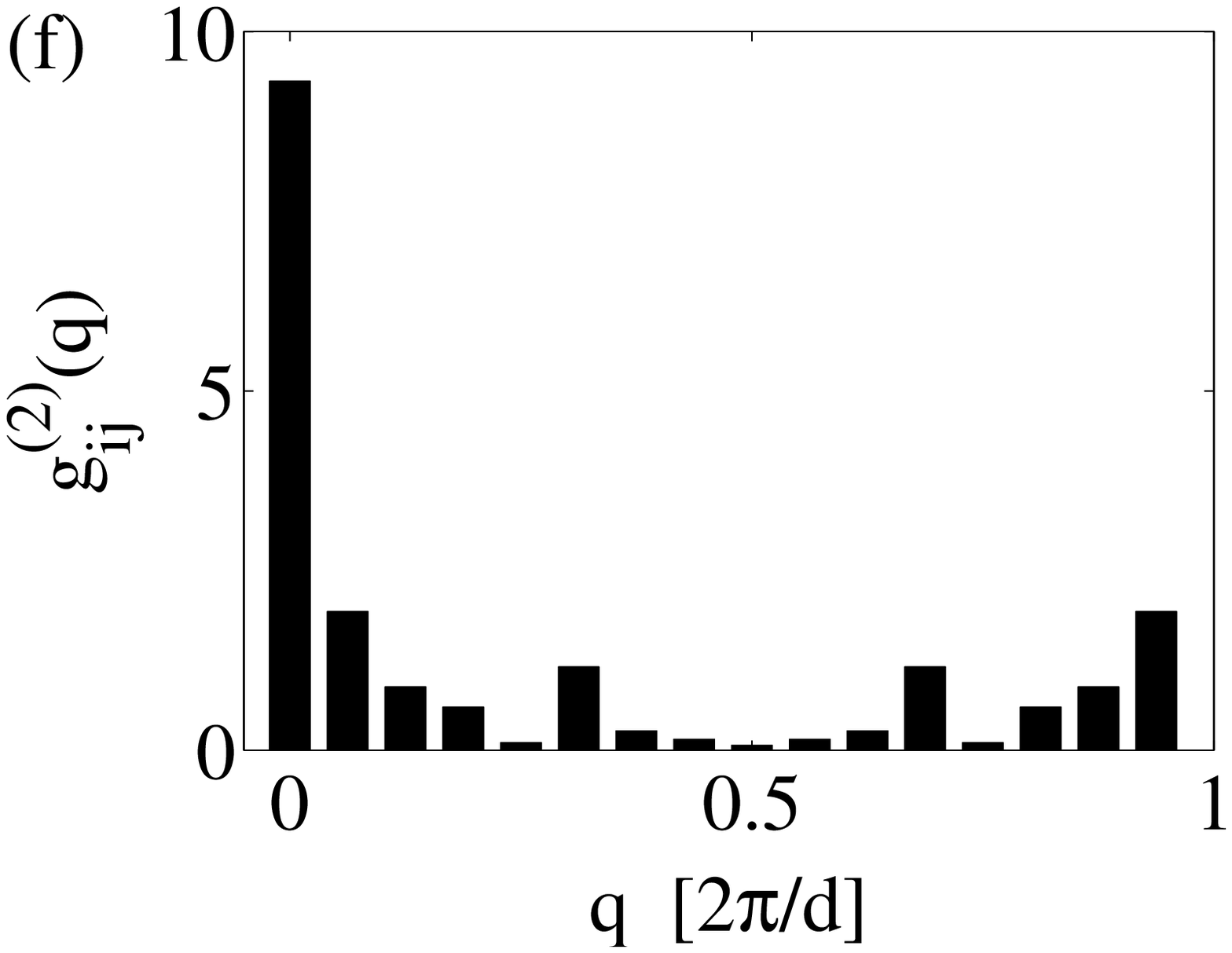}
  \end{minipage}
\caption{Ground state long range order [Eq.~\eqref{g2}] in position ($\mathbf{x}$) and quasi-momentum space ($q$) for two particles in a $16 \times 16$  lattice. The ground states are calculated for $\alpha = 0$ [(a) and (b)], $\alpha=1/2$ [(c) and (d)], and $\alpha =1/3$ [(e) and (f)]. One of the two points $\mathbf{x}_i$ is fixed close to the center of rotation while $\mathbf{x}_j$ is moved from end to the other parallel to one of the edges of the lattice. The asymmetry in (a),(c) and (e) is due to $\mathbf{x}_i$ being closer to one of the infinite walls.  \label{LongRangeOrder}}
\end{center}
\end{figure}

A useful tool for understanding the structure of a many-body state is the long range order, which, for a lattice system, can be defined by
\begin{equation}
g^{(2)}_{ij}(\mathbf{x}_i-\mathbf{x}_j)=\frac{\langle \hat{a}^{\dagger}_j \hat{a}^{\dagger}_i \hat{a}_i \hat{a}_j \rangle} {\langle \hat{a}^{\dagger}_j \hat{a}_j \rangle\langle \hat{a}^{\dagger}_i \hat{a}_i \rangle}\,. \label{g2}
\end{equation} 
 The hardcore nature ($U \rightarrow \infty$) of the particles manifests itself as an anti-correlation envelope in the ground state  seen most clearly for $\alpha=0$ [Fig.~\ref{LongRangeOrder}(a)]. The periodicity in the Hamiltonian for rational $\alpha$ discussed earlier is seen in the long range order of the ground state plotted for $\alpha=1/2$ and $\alpha=1/3$ [Figs.~\ref{LongRangeOrder}(c) and \ref{LongRangeOrder}(e)].  The Fourier transforms of the long range order [Figs.~\ref{LongRangeOrder}(d) and \ref{LongRangeOrder}(f)] show peaks at 
 $q=1/2d$ and $q=1/3d$ corresponding to periodicities of two and three lattice sites respectively. Interestingly, every second site is correlated for $\alpha=1/2$ while every third site is anti-correlated for $\alpha=1/3$.

\begin{figure}[t]
\begin{center}
   \includegraphics[width=7.8cm]{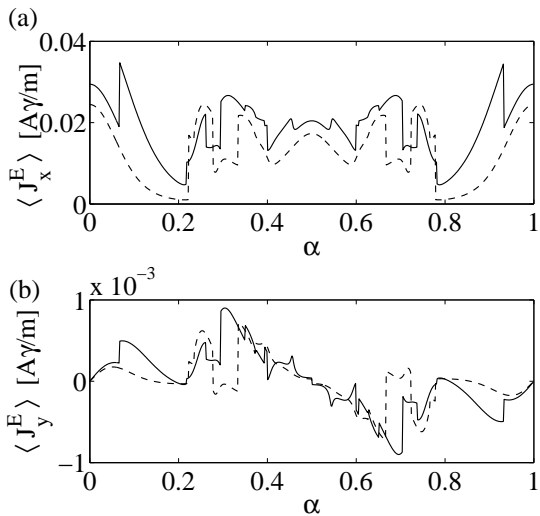}
   \caption{ End current response per particle for one (dashed, $n=1$) and two (solid, $n=2$) particles in a $8\times8$ lattice along the (a) longitudinal direction --- $\langle \hat{J}^E_x \rangle/n$, and (b) along the transverse direction --- $\langle\hat{J}^E_y \rangle /n$. The perturbation is modulated at a frequency $\nu = E_R/\hbar$.\label{EndCurrentsManyParticles}}
\end{center}
\end{figure}

The features of the end current response [Fig.~\ref{EndCurrentsSchematic}] as a function of $\alpha$ for two particles in a $8\times8 $ lattice are altered considerably due to finite-size effects [Fig.~\ref{EndCurrentsManyParticles}]. The longitudinal $\left( \langle J^E_x \rangle\right)$ and transverse $\left( \langle J^E_y \rangle\right)$ end currents display similar features close to $\alpha=1/2,1/3,2/3,1/4,3/4 \ldots$ though the transverse end currents are antisymmetric about $\alpha=0.5$. Both single-particle and two-particle end current responses are similar with a few additional peaks in the latter. In the single-particle analysis, the first distinct peaks ($\alpha=1/2,1/3,2/3$) start emerging for lattice sizes $>10\times 10$, a size just beyond our numerical methods for two particles. 

\begin{figure}[t]
\begin{center}
   \includegraphics[width=7.8cm]{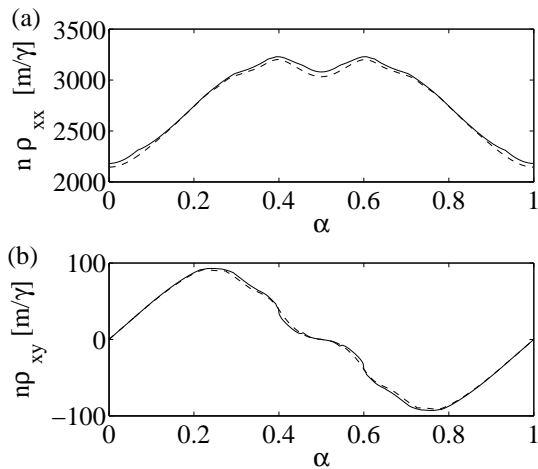}
   \caption{ Transverse and diagonal resistivity scaled by the number of particles $n$ as a function of $\alpha$ for one (dashed) and two (solid) particles in a $8\times8$ lattice. In the low filling (particles per lattice site) limit, the effect of the interaction is to decrease the conductivity per particle. The perturbation is modulated at a frequency $\nu = E_R/\hbar$. \label{ResistivityManyParticles}}
\end{center}
\end{figure}

The resistivity [Eq.~\eqref{rho}] scaled by the number of particles as a function of $\alpha$ is plotted in Fig.~\ref{ResistivityManyParticles}.  In the low filling limit (particles/site $< 0.1$), the effect of the interaction between particles is to enhance the scaled resistivity or, equivalently, lower the sample-averaged conductivity per particle. This is consistent with earlier findings~\cite{Bhat:2006a}, where increasing interaction reduced the current per particle. The two particle resistivity also shows weak dips (inflections) in the longitudinal (transverse) resistivity at fractional values of $\alpha$. Note that this calculation is in the very dilute limit, where the physics is largely dominated by single-particle effects.  

\begin{figure}[t]
\begin{center}
   \includegraphics[width=7.8cm]{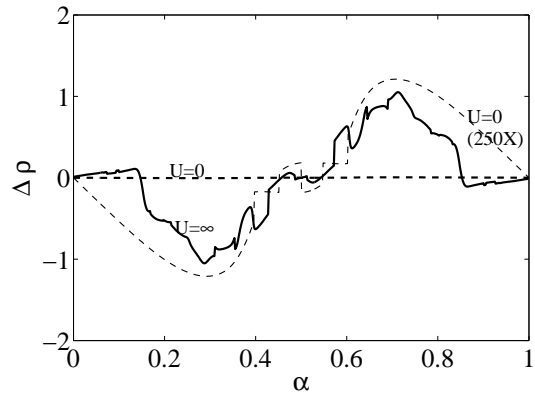}
   \caption{ Differential density response  for  four particles in a $4\times 4$ lattice in the strongly repulsive (solid) and non-interacting (dashed) limits. The two dashed lines describe the same limit but are displayed at different magnifications. In the strongly interacting limit, the redistribution due to the Coriolis force for this particular system is three orders of magnitude greater than that in the non-interacting limit. \label{DensityResponse}}
\end{center}
\end{figure}

One feature of the Hall effect is the breaking of number-density symmetry along the $y$--axis despite the perturbation being along the $x$--axis. This leads to charge buildup and eventually creates a stopping Hall potential. The number density asymmetry can be quantified by the difference in the number of particles on each half, where the lattice is divided in two along the direction of the perturbation. This quantity $\Delta \rho$ is plotted as a function of $\alpha$ for four particles in a $4\times 4$ lattice in Fig.~\ref{DensityResponse}. The effect of interaction on $\Delta \rho$ is examined by comparing the weak $(U=0)$ and strong $(U=\infty)$ interaction limits. In the strongly interacting regime, the redistribution (as quantified by $\Delta \rho$) due to the Coriolis force is three orders of magnitude greater than that in the non-interacting limit. The change in the direction of particle pileup at $\alpha=0.5$ marks the change in the direction of the Coriolis force.  Since the particles are charge neutral in this system, the retarding potential  along the transverse direction is created by strong repulsive interaction between particles.

\section{Conclusion}
This paper considers bosons in a rotating optical lattice that have a Hamiltonian similar to that for Bloch electrons in a magnetic field. The Hall effect in this system is probed using linear-response theory and the Kubo formula. The single-particle case exhibits fractional quantum Hall features. Density redistribution in small, strongly-correlated many-particle systems shows the equivalent of the classical effect and the mapping between the Coriolis and Lorentz forces. However, larger many-particle systems need to be considered to find the FQHE as described by dips (plateaus) in the diagonal (transverse) elements of the conductivity tensor as a function of $\alpha$. Tilting the rotating optical lattice in experiments such as \cite{Tung:2006} opens up this topic in two different ways. In the weakly interacting limit, the system can be probed to study the Coriolis force in superfluid systems. The system described by Tung {\it et.~al.}~\cite{Tung:2006} obtains an ensemble of stacked, identical two-dimensional layers. If the layers are stacked along the $z$--axis and the lattice is tilted along the $x$--axis, the effect of the Coriolis force would be to break the two-fold symmetry in the density distribution along the $y$--axis.  An observable to characterize this is obtained by imaging the density distribution along the $z$--axis using a CCD camera and evaluating $\Delta \rho$, the change in density induced by the perturbation. The strongly-interacting limit can be achieved by tuning the two-body scattering length using a Feshbach resonance. In this limit, the ground state  can be examined via Bragg scattering. Future work will focus on extending results obtained here to larger systems by improving the treatment of boundary conditions. This will be done using a combination of periodic boundary conditions and blocking the lattice into 'supercells'. 

\section{Acknowledgments}

We thank Jami Kinnunen, Dominic Meiser, Brandon M. Peden, Ronald A. Pepino, Brian T. Seaman, Volker Schweikhard, Shih-Kuang Tung and Jochen Wachter for several useful discussions. The authors acknowledge funding from the US Department of Energy, Office of Basic Energy Sciences via the Chemical Sciences, Geosciences, and Biosciences Division (R.~B.), Deutsche Forschungsgemeinschaft (M.~K.), and the National Science Foundation (M.~K., J.~C., M.~J.~H).

\end{document}